\documentclass[11pt,english,longbibliography]{article}
\usepackage{url}

\usepackage{draft} 
\usepackage{putex}

\usepackage{hyperref}

\renewcommand{\Re}{{\rm Re \, }}

\newcommand{\CO}{{\cal O}}

\newcommand{\CF}{{\cal F}}

\newcommand{\CI}{{\cal I}}

\newcommand{\SO}{{\text{SO}}}

\newcommand{\CJ}{{\cal J}}

\usepackage{amsthm}

\makeatletter
\newcommand*{\rom}[1]{\expandafter\@slowromancap\romannumeral #1@}
\makeatother

\usepackage[latin9]{inputenc}

\usepackage{verbatim}
\usepackage{prettyref}
\usepackage[numbers,sort&compress]{natbib}
\usepackage{amsmath}
\usepackage{amssymb}
\usepackage{tikz-cd}
\tikzset{commutative diagrams/row sep/huge=4cm}
\tikzset{commutative diagrams/column sep/huge=4cm}
\tikzcdset{scale cd/.style={every label/.append style={scale=#1},
    cells={nodes={scale=#1}}}}
\usepackage{lmodern}
\usepackage{tcolorbox}
\usepackage{cases}
\usepackage{bbold}
\usepackage{bm}
\usepackage{ytableau}
\usepackage{youngtab}
\usepackage{mathtools}
\usepackage{graphicx}
\usepackage[nottoc]{tocbibind}
\usepackage{mathrsfs}
\usepackage{caption}
\usepackage{subcaption}
\usepackage{cancel}
\usepackage{float}
\usepackage{tikz}

\usetikzlibrary{arrows.meta}
\usetikzlibrary{bending}
\usepackage{tikz}

\tikzset{
    Witten diagram/.style={
        execute at begin picture={
            \draw[blue, line width=1.5pt] circle[radius=\pgfkeysvalueof{/tikz/Witten/radius}];
            \path node (X){\phantom{X}};
        },
        baseline={(X.base)}
    },
    vertex/.style={circle,fill,inner sep=1.5pt,node contents={}},
    Witten/.cd,
    radius/.initial=3cm
}
\usetikzlibrary{patterns}
\usepackage{lipsum}
\usepackage{framed}
\usepackage{adjustbox}
\usepackage{mathtools}
\usepackage{braket}

 \usepackage{slashed}
\usepackage{esvect}
\usepackage{dsfont}

\usepackage{color}
\definecolor{darkgreen}{rgb}{0,0.5,0}
\definecolor{darkblue}{rgb}{0,0,0.6}
\definecolor{purple}{rgb}{0.4,.2,0.7}

\numberwithin{equation}{section}
\numberwithin{figure}{section}
\numberwithin{table}{section}

\def\CG{{\cal G}} 
\def\CQ{{\cal Q}}

\def\CN{{\cal N}}

\newcommand{\ep}{\epsilon}

\newcommand{\CB}{\mathcal{B}}
\newcommand{\CA}{\mathcal{A}}

\DeclareFontShape{OT1}{cmr}{mx}{n}{<->cmr10}{}

\begin{document}

\preprint{PUPT-2655}

\title{\centering Surprises in the Ordinary:
\\
$O(N)$ Invariant Surface Defect in the $\epsilon$-expansion
}

\authors{Oleksandr Diatlyk,\worksat{\NYU} Zimo Sun,\worksat{\PUJ,\PUG} and Yifan Wang\worksat{\NYU}}

\institution{NYU}{Center for Cosmology and Particle Physics, New York University, New York, NY 10003, USA}

\institution{PUJ}{Joseph Henry Laboratories, Princeton University, Princeton, NJ 08544, USA}

\institution{PUG}{Princeton Gravity Initiative, Princeton University, Princeton, NJ 08544, USA}


\abstract{We study an $O(N)$ invariant surface defect in the Wilson-Fisher conformal field theory (CFT) in $d=4-\epsilon$ dimensions. This defect is defined by mass deformation on a two-dimensional surface that generates localized disorder and is conjectured to factorize into a pair of ordinary boundary conditions in $d=3$. We determine defect CFT data associated with the lightest $O(N)$ singlet and vector operators up to the third order in the $\epsilon$-expansion, find agreements with results from numerical methods and provide support for the factorization proposal in $d=3$. Along the way, we observe surprising non-renormalization properties for surface anomalous dimensions and operator-product-expansion coefficients in the $\epsilon$-expansion. We also analyze the full conformal anomalies for the surface defect.}

\date{}

\maketitle

\tableofcontents

\section{Introduction and Summary}

Defects play an active role in the modern understanding of Quantum Field Theory (QFT) (see \cite{Andrei:2018die} for a review). They are natural generalizations of the familiar point-like local operators and are defined on higher-dimensional submanifolds of the spacetime such as lines and surfaces. These defect operators encode much richer features of the underlying theory, intuitively by introducing non-local probes to the system, and thus are sensitive to phase structure that is ambiguous or harder to access by point-like observables. Defects also harbor interesting dynamics on their worldvolume, as a result of nontrivial interactions among degrees of freedom on the defect and in the bulk. This is particularly the case when the bulk is gapless and described by a Conformal Field Theory (CFT). In this case, universality classes of coupled bulk-defect systems are governed by conformal defects or defect CFT (DCFT) \cite{Billo:2016cpy}. Due to the enhanced conformal symmetry, this provides a fruitful playground to explore and analyze defect dynamics. 

Our knowledge of defect operators (and DCFTs) is still primitive, partly because they 
do not follow the same set of rules as the point-like local operators.
To this end, it is indispensable to come up with tractable yet rich examples of defects to aid us in the exploration and eventually towards a complete formulation in CFT and general QFT that incorporates such non-local observables.

In this paper, we study a simple conformal surface defect with interesting dynamics in the $O(N)$ Wilson-Fisher (WF) CFT that was recently introduced and analyzed in \cite{Krishnan:2023cff,Trepanier:2023tvb,Raviv-Moshe:2023yvq,Giombi:2023dqs} using large $N$ and $\epsilon$-expansion methods (a similar and related defect of codimension-one was studied long time ago in \cite{Bray:1977fvl}). We refer to this defect as the ordinary surface defect in the WF CFT (the terminology will be clear shortly). The WF CFT is arguably the simplest interacting CFT: it is a field theory of $N$ scalar fields $\phi^I$ with $I=1,2,\dots,N$ interacting through an $O(N)$ invariant quartic potential \cite{Zinn-Justin:1989rgp}.  The surface defect of interest has a simple definition in the UV (at short distance) by the integrated insertion of the mass operator $(\phi^I)^2$ in the $O(N)$ CFT on $\mathbb{R}^2$ in the spacetime. For $d<4$, this mass operator has dimension $\Delta<2$ and when turned on in the bulk, it is responsible for the well-known order-disorder phase transition in the $O(N)$ model. If instead this mass deformation is concentrated on a surface, it creates local order or disorder (depending on the sign) on this surface,
and triggers a defect renormalization group (RG) flow from the trivial (transparent) surface defect to a nontrivial conformal surface defect. In particular, for $d=3$, this surface defect becomes an interface. Results from \cite{Bray:1977fvl,Krishnan:2023cff} in the large $N$ limit suggest that for one sign of the mass deformation (that creates disorder), the conformal interface in the IR is ${\cal I}_+=|{\rm Ord}\ra \la {\rm Ord}|$ and factorized into a pair of $O(N)$ invariant conformal boundary conditions of the CFT in the ordinary universality class denoted by $|{\rm Ord}\ra$. This is why we refer to this surface universality class (for general $d<4$) as the ``ordinary'' type. For the other sign of the mass deformation (that creates order), the defect is of the ``extraordinary-log'' type. Its IR description involves a factorized interface  ${\cal I}_-=|\hat n \ra \la \hat n |$ from a pair of boundaries of the normal universality class denoted by $|\hat n\ra$ (which breaks the $O(N)$ symmetry to $O(N-1)$ via a unit $N$-vector $\hat n$), coupled to local order ($i.e.$ 2d Goldstone bosons, also denoted by $\hat n$ in \cite{Krishnan:2023cff}) to restore the $O(N)$ symmetry.\footnote{This coupling is nontrivial for $N\geq 2$ and marginally irrelevant thus experiences logarithmical running \cite{Krishnan:2023cff}. For $N=1$, namely the Ising CFT, we expect this interface to be factorized and decomposable in the IR, given by $|+\ra \la +|\oplus |-\ra\la -|$, where $|\pm \ra$ denote the two normal boundary universality classes of the Ising CFT. }  Compared to the interface (or the surface in general $d<4$), the boundary conditions of the $O(N)$ model have been studied much more extensively \cite{Diehl:1996kd}. In particular, the recent fuzzy sphere technique \cite{Zhu:2022gjc} for solving CFT has been applied to extract a large variety of conformal data for the ordinary and normal boundary universality classes for the Ising CFT in $d=3$ \cite{Dedushenko:2024nwi,Zhou:2024dbt}.

A main goal of this paper to test this IR factorization property of the ordinary conformal surface defect at finite $N$ in $d=3$
using $\epsilon$-expansion and its Pad\'e  resummation  from $d=4-\epsilon$. Along the way, we 
push existing perturbative results from $\epsilon$-expansion to the next order, which are needed to carry out the Pad\'e  resummations (we will comment on other resummation techniques shortly). See
Table~\ref{Dphicomp} for an example of our results for the anomalous dimensions on the surface defect and comparison with Monte Carlo results for boundaries in $d=3$. Furthermore, 
we 
also extract new conformal data beyond critical exponents associated with this surface defect, such as one- and two-point functions of bulk and defect operators made of the fundamental scalar field $\phi^I$, which are among the basic operator-product-expansion (OPE) coefficients that determine general conformal observables in the DCFT and obey nontrivial constraints from bootstrap equations \cite{Bianchi:2015liz} that we demonstrate explicitly. Along the way, we discover surprising non-renormalization properties associated with these DCFT data. For example, we find that the UV and IR scaling dimensions of the mass operator $(\phi^I)^2$ on the surface satisfy a version of the ``shadow relation'' to the second order in the $\epsilon$-expansion (see around \eqref{surp} for the precise expression)\footnote{This is somewhat reminiscent of the shadow relations found previously for other observables in the $O(N)$ CFT and its long-range generalizations
\cite{Paulos:2015jfa,Behan:2017emf,Goncalves:2018nlv,Alday:2019clp,Giombi:2019enr}.} 
\ie
\Delta^{\rm UV}_{(\phi^I )^2} + \Delta^{\rm IR}_{(\phi^I )^2} \approx 4\,.
\label{shadowrel}
\fe
While the above relation receives nonzero corrections at the $\epsilon^3$ order, interestingly, by comparing to numerical results for both scaling exponents from conformal bootstrap and Monte Carlo at $d=3$ (together with the assumption of factorization for the interface), we find surprisingly good agreements with this approximate shadow relation in \eqref{shadowrel}. 
We also find similar non-renormalization properties for the correlator of the fundamental scalar with its cousin on the ordinary surface (see around \eqref{surp2}). 

Conformal defects admit richer conformal anomalies compared to CFTs. We study conformal anomalies universal to surface defects in dimension $d\geq 3$, which produce the following anomalous trace of the canonical conformal stress energy tensor $T_{\m\n}$,\footnote{For conformal surface defects in dimension $d\geq 4$, another anomaly is possible and is proportional to the pullback of the bulk Weyl curvature tensor to the surface $\Sigma$.}
\ie 
\la T^\m_\m \ra = {1\over 24\pi}\D(\Sigma)\left (b  R_\Sigma + d_1 \left (K^a_{ij} K_a^{ij}- {1\over 2} K^a K_a \right) \right)
\label{confanom}
\fe
localized to the defect worldvolume $\Sigma$ \cite{Graham:1999pm,Henningson:1999xi,Schwimmer:2008yh}. Here $R_\Sigma$ is the Ricci curvature scalar on $\Sigma$, $K^a_{ij}$ is the extrinsic curvature tensor with $a$ and $i,j$ labeling the directions normal and tangential to $\Sigma$ respectively, and $K^a$ is its trace. The conformal anomaly coefficients are $b$ and $d_1$ in \eqref{confanom}. The $b$ anomaly is a direct generalization of the familiar $c$ anomaly for 2d CFT and obeys monotonicity under defect RG flows \cite{Jensen:2015swa,Casini:2018nym,Kobayashi:2018lil,Wang:2020xkc,Shachar:2022fqk,Casini:2023kyj}. It encodes the logarithmic divergence in the expectation value of the surface defect ($e.g.$ on a sphere).
On the other hand, the $d_1$ anomaly is novel to the defect and does not have monotonic properties under defect RG in general. It is known to satisfy the relation \cite{Bianchi:2015liz},
\ie 
d_1={ 3\pi^2\over 4}C_{\rm D}\,,
\label{d1CD}
\fe
where $C_{\rm D}$ is the Zamolodchikov norm that determines the two-point function of the canonically defined displacement operator ${\rm D}^a$ (which results from the broken transverse translation symmetries due to the defect insertion) for flat surface defect in $\mR^d$  \cite{Billo:2016cpy},
\ie 
\la {\rm D}^a(x) {\rm D}^b(0) \ra=C_{\rm D}\D^{ab}{1\over |x|^6}\,.
\fe
In this paper, we calculate both the $b$ and the $d_1$ conformal anomalies for the ordinary surface defect in the WF CFT. We provide the first analytic result for the $d_1$ anomaly of a non-supersymmetric conformal surface defect in an interacting CFT at finite $N$.\footnote{See also previous works on this defect anomaly in free theory \cite{Jensen:2015swa} and large $N$ CFT \cite{Herzog:2020lel} as well as its relation to entanglement entropy \cite{Bianchi:2015liz}.} We  also find agreements with previous results for $b$ and extend them to the next order in the $\epsilon$-expansion. By extrapolating to $\epsilon=1,2$ where the DCFT is unitary, our results are consistent with the $b$-theorem.

There are a number of interesting future directions both on the analytic and on the numerical fronts towards improving our understanding of surface defects. 
Firstly we would like to understand the origin of the non-renormalization properties we observe for the ordinary surface defect: if relations such as \eqref{shadowrel} hold more generally (even only approximately), they can provide important insights into the defect dynamics. One potentially fruitful strategy is to apply a generalization of the analytic conformal bootstrap method 
(see \cite{Bissi:2022mrs} for a recent review) which has been used to solve and elucidate structures in the $\epsilon$-expansion of bulk CFT data \cite{Rychkov:2015naa,Gopakumar:2016wkt,Gopakumar:2016cpb,Liendo:2017wsn,Gliozzi:2017gzh,Alday:2017zzv,Henriksson:2018myn,Gopakumar:2018xqi,Guha:2019ipe,Carmi:2020ekr} (see also 
\cite{Bissi:2018mcq,Ferrero:2019luz,Dey:2020jlc,Bertucci:2022ptt,Bianchi:2022ppi,Gimenez-Grau:2022ebb,Bianchi:2022sbz,Herzog:2022jlx,Nishioka:2022qmj,Nishioka:2022odm,Dey:2024ilw} for recent related works on DCFT data using this method). Going beyond leading perturbative results,
it is well-known that the $\epsilon$-expansion of CFT data in the $O(N)$ model is asymptotic \cite{Brezin:1976vw} and various resummation techniques, including Pad\'e approximants and Borel transforms (see \cite{Zinn-Justin:1999opn} for a summary), have been developed to produce convergent results that match better with direct numerical analysis in $d=3$ ($i.e$ $\epsilon=1$) using Monte Carlo and conformal bootstrap techniques (see recent works \cite{Henriksson:2022rnm,Bonanno:2022ztf} and references therein).
Although the resummation introduces theoretical ambiguities (depending on the specific procedure and parameters in each procedure), there is strong evidence that the large order behavior of the anomalous scaling dimensions of bulk local operators in $\epsilon$-expansion is governed by instantons, thus indicating Borel summability and suggesting the Borel approach to analyze the series (possibly in combination with other techniques, such as Borel-Pad\'e) \cite{Lipatov:1976ny,Zinn-Justin:1989rgp,Dunne:2021lie}. 
It would be very interesting to develop a better understanding of the nature of the $\epsilon$-expansion for defect observables in the $O(N)$ CFT, such as the large order behavior for anomalous dimensions on the defect, 
which would inform us the more natural resummation technique to implement in order to extract physical DCFT data at integer dimensions.
In a different direction,
as shown recently by studying the fusion of defects \cite{Soderberg:2021kne,Rodriguez-Gomez:2022gbz,SoderbergRousu:2023zyj,Diatlyk:2024zkk} and the effective field theory that emerges in the fusion limit \cite{Diatlyk:2024qpr,Kravchuk:2024qoh,Cuomo:2024psk}, one can deduce universal DCFT data, such as the asymptotic behavior of bulk one-point functions with the defect. It would be interesting to carry out this analysis explicitly for the ordinary surface defect in the WF CFT. Finally it would be worthwhile to pursue a direct analysis of the ordinary interface defect in the $d=3$ Ising CFT and more general $O(N)$ CFTs using the fuzzy sphere method \cite{Zhu:2022gjc,Dedushenko:2024nwi,Zhou:2024dbt}, to test the factorization proposal in \cite{Bray:1977fvl,Krishnan:2023cff} at finite $N$.

The rest of the paper is organized as follows. We start with a brief review of the ordinary surface defect in the WF CFT in Section~\ref{nac}. In Section~\ref{MassOperator}, we study the OPE of the bulk mass operator, which is the lightest $O(N)$ singlet operator, with the surface defect and the DCFT data therein. In Section~\ref{sec:fund}, we focus on the fundamental scalar field $\phi^I$, which is lightest among all nontrivial operators, and its leading OPE with the ordinary surface defect. In Section~\ref{disO}, we analyze displacement operators for the ordinary surface defect, determine the Zamolodchikov norm, and confirm explicitly the fundamental Ward identity involving the displacement operator. Finally in Section~\ref{bcal}, we compute the $b$ anomaly for the ordinary surface defect and verify the $b$-theorem in this setting. Technical details in the perturbative calculations have been relegated to the appendices.

\section{Review of the $O(N)$ Invariant Surface Defect}\label{nac}
The action of the massless $O(N)$ symmetric Wilson-Fisher (WF) model in $d=4-\epsilon$ dimensions reads 
\begin{align}\label{SON}
    S_{O(N)}=\int d^{d}X \left[\frac{1}{2}\left(\partial\phi_0^I\right)^2+\frac{\lambda_0}{4!}(\phi_0^I\phi_0^I)^2\right]\,,
\end{align}
where the superscript $I=1,2,\cdots, N$ is an $O(N)$ vector index and the subscript ``0'' indicates bare quantities. The spacetime coordinates are denoted by $X^\m$.
When $\lambda_0=0$, the propagator for a fixed index $I$ is given by
\begin{align}
G_d (X_1, X_2)= \langle \phi_0^I(X_1)\phi_0^I(X_2)\rangle_{0}=  \frac{C_\phi}{|X_{12}|^{d-2}}\,, \quad C_\phi\equiv \frac{\Gamma(\frac{d}{2}-1)}{4\pi^{\frac{d}{2}}}\,.
\end{align} 
We denote the renormalized coupling by $\lambda$. Up to $\CO(\lambda^4)$, the bare coupling $\lambda_0$ and the beta function of $\lambda$ are \cite{Kleinert:2001ax},\footnote{In the main text, we mostly focus on DCFT data that can be extracted from a three-loop analysis of the bulk-defect system ($i.e.$ we determine the bare coupling $\lambda_0$ and its beta function up to $\lambda^3$ as well as the wavefunction renormalization $Z_{\phi^2}$ \eqref{wavefunctionrenormorderl3} to order $\lambda^2$).  In Appendix \ref{ordereps3}, we extend the analysis to four-loop and determine certain DCFT data to order $\epsilon^3$, incorporating contributions up to $\lambda^4$.}

\begin{align}\label{lambdabare}
    \lambda_0 &=\mu^{\epsilon} \bigg [ \lambda +   \frac{ N+8}{3 \epsilon} \frac{\lambda^2}{(4\pi)^2}
+  \left( \frac{\left( N+8\right)^2}{9 \epsilon^{2} }  -  \frac{3N+14 }{6 \epsilon}  \right) \frac{\lambda^3}{(4\pi)^4}\\  
&
+  \bigg( \frac{\left(N+8\right)^3}{27\epsilon^{3}}   -  \frac{7 \left(N+8\right) \left( 3N+14\right)}{54 \epsilon^{2}}  
+ \frac{2960 + 922N + 33N^2 + 96 \zeta(3)\left(5N+22\right)}{648 \epsilon}   \bigg)\frac{\lambda^4}{(4\pi)^6}
\bigg]\,,\nonumber
\end{align}
and
\begin{gather}
    \beta_\lambda=- \epsilon\lambda  +\frac{(N\!+\!8)\lambda^2}{3(4\pi)^2}-\frac{(3N\!+\!14)\lambda^3}{3(4\pi) ^4}+\frac{33N^2\!+\!922 N\!+\!2960\!+\!96 \zeta(3)(5N\!+\!22)}{216}\frac{\lambda^4}{(4\pi)^6}\,. 
    \label{betalambda}
\end{gather}
The fixed point then follows,\footnote{The order $\ep^3$ piece for the bulk fixed point $\lambda_\star$ is included here for completeness. It is not necessary in the three-loop analysis but will be important when we determine the order $\ep^3$ corrections to the DCFT data in the four-loop analysis (see Appendix~\ref{ordereps3}).}

\begin{align} \label{fixedlambda}
    \frac{\lambda_\star}{(4\pi)^2}= &\,\frac{3\, \epsilon }{N+8}+ (3 N+14)\frac{ 9\,\epsilon ^2}{(N+8)^3} \\
    +& \left[ \frac{3}{8} \left( - 33N^3+110N^2+1760N+4544 \right) - 36 \zeta(3) \left( N+8\right) \left( 5N+22\right) \right]\frac{\epsilon^3}{( N+8)^5} \,.  \nonumber
\end{align}

Following \cite{Trepanier:2023tvb,Raviv-Moshe:2023yvq,Giombi:2023dqs}, we introduce a surface defect in the $O(N)$ model by adding 
\begin{align}\label{SDdef}
    S_D \equiv  h_0\int_\Sigma d^2 X\, \phi_0^2\,, \quad \phi_0^2\equiv \phi_0^I \phi_0^I
\end{align}
to $S_{O(N)}$, where $\Sigma$ is a two-dimensional submanifold in $\mathbb{R}^d$ and  $d^2 X$ denotes the induced measure on the defect. 
We will consider $\Sigma = \mathbb{R}^2$ in most part of the paper and $\Sigma = S^2$ in Section \ref{bcal}.
When $\Sigma = \mathbb{R}^2$, we place it in the $X^1\!-\!X^2$ plane, with $x^i \equiv  (X^1, X^2)$ being coordinates on the defect  and $\eta^a \equiv (X^3, X^4, \cdots, X^d)$ being the  transverse coordinates. In this case, the induced measure is $d^2\sigma= dx^1 dx^2\equiv d^2 x$.
When $\Sigma = S^2$, we take the sphere to be at $(X^1)^2+(X^2)^2+(X^3)^2 =1$ and $X^{\m>3}=0$.\footnote{The radius of the sphere is set to be 1 here for convenience. We will restore the radius $R$ when computing the $b$ function in Section \ref{bcal}.} We introduce stereographic coordinates $y^i$ on the $S^2$ so that the measure is $d^2\sigma=\Omega(y)^2 d^2 y$, where $\Omega (y) = \frac{2}{1+y^2}$. Given two points $X_1=X(y_1)$ and $X_2=X (y_2)$ on the $S^2$, the $\SO(d+1)$ invariant distance is 
\begin{align}\label{invariantdistnce}
|X_{12}| \equiv \sqrt{\left(X(y_1)-X(y_2)\right)^2} = \sqrt{\Omega(y_1)\Omega (y_2)}\, |y_{12}|\,.
\end{align}
Under an inversion $y^i\to y^i/y^2$, we have the following transformation laws
\begin{align}\label{inversion}
\Omega(y) y^2\to \Omega(y)\,, \quad  \Omega(y_1)\Omega (y_2) y_{12}^2\to  \Omega(y_1)\Omega (y_2) y_{12}^2\,, \quad \Omega(y)^2 d^2 y\to \Omega(y)^2 d^2 y\,.
\end{align}
These transformation laws are very useful when we evaluate integrals on spheres.

\section{Mass Operator in the Bulk and on the Defect}\label{MassOperator}

\begin{figure}[!htb]
\centering
\begin{tikzpicture} 
\draw [line width = 0.5mm] (-1.2, -1) to (1.2,-1);
\draw   (0,1)node[vertex]{} to  (-1,-1)node[vertex]{}  to [out=45,in=135]  (0,-1)node[vertex]{}  to  [out=45,in=135] (1,-1)node[vertex]{} to (0,1);
\node at (0.1,-1.5) {$\CJ_{0,3}$};
\end{tikzpicture}
\qquad 
\begin{tikzpicture} 
\draw [line width = 0.5mm] (-1.2, -1) to (1.2,-1);
\draw   (0,1) node[vertex]{} to[out=-45,in=45]  (0,0)node[vertex]{} to[out=135,in=-135]  (0,1);
\draw   (0,0) to (-1,-1)node[vertex]{} to [out=45,in=135] (1,-1)node[vertex]{} to (0,0)  ; 
\node at (0.1,-1.5) {$\CJ^{(1)}_{1,2}$};
\end{tikzpicture}
\qquad 
\begin{tikzpicture} 
\draw [line width = 0.5mm] (-1.2, -1) to (1.2,-1);
\draw   (0,1) node[vertex]{} to (-0.5,0)node[vertex]{} to[out=-140,in=80]  (-1,-1)node[vertex]{};
\draw  (-0.5,0) to[out=-100,in=40]  (-1,-1)node[vertex]{};
\draw  (-0.5,0)   to (1,-1);
\draw   (0,1) to (1,-1) node[vertex]{}; 
\node at (0.1,-1.5) {$\CJ^{(2)}_{1,2}$};
\end{tikzpicture}
\\
\begin{tikzpicture} 
\draw [line width = 0.5mm] (-1.2, -1) to (1.2,-1);
\draw   (-1,0) node[vertex]{} to[out=15,in=165]  (1,0)node[vertex]{};
\draw   (-1,0) to[out=-15,in=-165]  (1,0);
\draw   (0,1) node[vertex]{} to  (-1,0) to (0,-1) node[vertex]{} to (1,0) to (0,1);
\node at (0.1,-1.5) {$\CJ_{2,1}^{(1)}$};
\end{tikzpicture}
\qquad
\begin{tikzpicture} 
\draw [line width = 0.5mm] (-1.2, -1) to (1.2,-1);
\draw   (0,1) node[vertex]{} to[out=-45,in=45]  (0,0.3)node[vertex]{} to[out=-45,in=45]  (0,-0.3)node[vertex]{} to[out=-45,in=45]  (0,-1)node[vertex]{} to[out=135,in=-135]  (0,-0.3) to[out=135,in=-135]  (0,0.3) to[out=135,in=-135]  (0,1);
\node at (0.1,-1.5) {$\CJ_{2,1}^{(2)}$};
\end{tikzpicture}
\qquad
\begin{tikzpicture} 
\draw [line width = 0.5mm] (-1.2, -1) to (1.2,-1);
\draw   (0,1) node[vertex]{} to[out=-135,in=135]  (0,-1)node[vertex]{} to [out=45,in=-90] (0.38,-0.3) node[vertex]{} to (0.38,0.3) node[vertex]{} to [out=90,in=-45](0,1);
\node at (0.1,-1.5) {$\CJ_{2,1}^{(3)}$};
\draw  (0.38,-0.3) to [out=45,in=-45] (0.38,0.3)  [out=-135,in=135] to (0.38,-0.3); 
\end{tikzpicture}
\caption{Diagrams contributing to $\langle \phi^2(\eta)\rangle$ at order $h_0^3, \lambda_0 h_0^2$ and $\lambda_0^2 h_0$.}
\label{4diabeta}
\end{figure}
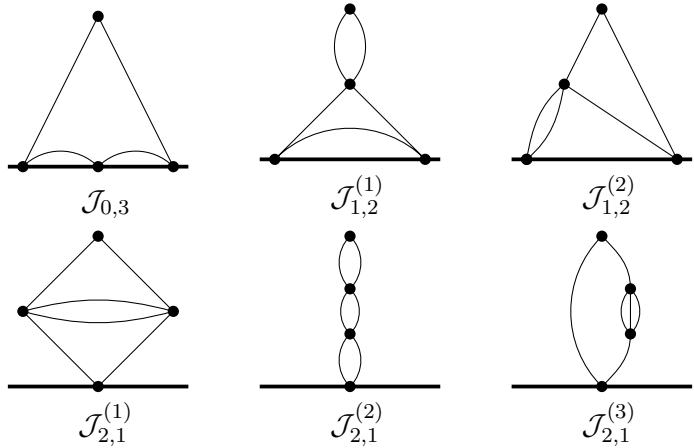
In this section, we carry out the renormalization of the defect coupling $h_0$ to the
third order in coupling constants, by computing the one-point function of the bulk $O(N)$ invariant operator $\phi_0^2\equiv\phi_0^I \phi_0^I$. Based on this result, we will also derive the scaling dimension of the defect operator $\hat\phi_0^2=\hat\phi_0^I \hat\phi_0^I$ which dominates the bulk-defect OPE beyond the identity operator.\footnote{Throughout this paper, we will denote operators on the defect with a hat in order to distinguish them from their bulk counterpart. The only exception is the displacement operator considered in Section \ref{disO}.} In these calculations, we will set the location of the bulk operator to be $X=(0,\eta)$ using the translation symmetry along the defect. We will work with the minimal subtraction (MS) scheme.

Up to second order in coupling constants, the one-point function of the bare quadratic operator $\phi_0^2$ was computed in \cite{Trepanier:2023tvb,Raviv-Moshe:2023yvq,Giombi:2023dqs}. Their results can be concisely written as
\ie 
\langle\phi_0^2(0, \eta)\rangle_{2-\text{order}} =A_0 +A_1+B\,,
\fe 
where
\begin{align}
       A_0&= -\frac{2 h_0 N\pi C_\phi^2}{(d-3)\eta^{2(d-3)}}\,,\nonumber\\
       A_1&=4N  \frac{h_0^2 C_\phi^3 \Gamma_0(2; \frac{d-2}{2},\frac{d-2}{2},\frac{d-2}{2})}{\eta^{3d-10}}\,,\nonumber\\
       B &= \frac{N(N+2)}{3}\frac{\pi\lambda_0 h_0 C_\phi^4 C_{d; d-2, d-2}}{(\frac{3d}{2}-5)\eta^{3d-10}}\,.
\end{align}
In the above, the function $\Gamma_0$ is defined by \eqref{G0} and $C_{d; d-2, d-2}$ is given by \eqref{Cdef}.

At third order in coupling constants, there are six diagrams as shown in Figure~\ref{4diabeta}. The corresponding position space integrals are slightly unconventional due to a mixture of integrals in the bulk and on the defect. Nonetheless, we can evaluate them systematically using the Mellin-Barnes method (see Appendix~\ref{AI}).
For illustration, we describe the calculation of the first two diagrams in detail.
For $\CJ_{0, 3}$, the relevant integral is 
\begin{align}
\CJ_{0, 3} = -8Nh_0^3 C_\phi^4\int \frac{d^2 x_1 d^2 x_2 d^2 x_3}{x_{12}^{d-2}x_{23}^{d-2}(x_1^2+\eta^2)^{\frac{d-2}{2}}(x_3^2+\eta^2)^{\frac{d-2}{2}}}\,.
\end{align}
With a simple rescaling, we pull out the overall $\eta$ dependence and  perform the $x_2$ integral using \eqref{Cdef}. The  remaining double integral over $x_1$ and $x_3$ is a special case of \eqref{G0int}. Altogether, we have 
\begin{align}
\CJ_{0, 3}& =-8NC_\phi^4\, C_{2; \frac{d-2}{2},\frac{d-2}{2}}\Gamma_0\,\left(2;\frac{d-2}{2},\frac{d-2}{2},d-3\right)\frac{h_0^3}{\eta^{4d-14}}\nonumber\\
   &=-\left[\frac{1+\epsilon  (3+2 \log (\pi e^{\gamma_E}) )}{2^6 \pi ^5 \epsilon ^2}+\CO(1)\right]\frac{8Nh_0^3}{\eta^{2(1-2\epsilon)}}\,,
\end{align}
where $\gamma_E$ is the Euler-Mascheroni constant.
The  diagram $\CJ^{(1)}_{1, 2} $ involves a bulk point, denoted by $X_0$, and two points on the defect, denoted by $X_1$ and $X_2$ respectively. The relevant integral is 
\begin{align}
\CJ^{(1)}_{1, 2} = -\frac{2N(N+2)}{3}\lambda_0 h_0^2 C_\phi^5\int \frac{d^2 X_1 d^2 X_2 d^d X_0}{X_{03}^{2(d-2)}X_{01}^{d-2}X_{02}^{d-2}X_{12}^{d-2}}\,,
\end{align}
where $X_1=(x_1, 0)$, $X_2=(x_2, 0)$ and $X_3 = (0, \eta)$. We integrate out the bulk point using \eqref{Sdef0} and the defect points using \eqref{G0int}. These steps lead to the following MB representation of $\CJ^{(1)}_{1, 2}$
\begin{align}
\CJ^{(1)}_{1, 2} &= -\frac{2N(N+2)}{3}\frac{C_\phi^5\lambda_0 h_0^2}{\eta^{4d-14}}\nonumber\\
&\times\int\limits_{-i\infty}^{+i\infty}\frac{dz_1dz_2}{(2\pi i)^2} S_d\left(\frac{d\!-\!2}{2},\frac{d\!-\!2}{2}, d\!-\!2; z_1, z_2 \right) \Gamma_0\left(2;-z_1, -z_2, 2d\!-\!5\!+\!z_1\!+\!z_2\right)\,,
\end{align}
from which we can extract the poles in its $\epsilon$-expansion using the \texttt{Mathematica} program \texttt{MB.m} \cite{Czakon:2005rk}
\begin{align}
\CJ^{(1)}_{1, 2} = -\left[\frac{ 1+\epsilon  (3+2 \log (\pi e^{\gamma_E}) )}{2^8 \pi ^6 \epsilon ^2}+\CO(1)\right]\frac{N(N+2)\lambda_0 h_0^2}{\eta^{2(1-2\epsilon)}}\,.
\end{align}
The third diagram in Figure \ref{4diabeta} can be evaluated similarly
\begin{align}
    \CJ^{(2)}_{1, 2} &=-\frac{4N(N+2)}{3}\frac{\lambda_0 h_0^2}{\eta^{4d-14}}C_{\phi}^5\nonumber\\
    &\times\int\limits_{-i\infty}^{+i\infty}\frac{dz_1dz_2}{(2\pi i)^2} S\left(d-2,\frac{d-2}{2},\frac{d-2}{2}; z_1, z_2 \right) \Gamma_0\left(2;-z_1, \frac{d-2}{2}-z_2, \frac{3}{2}d-4+z_1+z_2\right)
    \nonumber\\
    &= -\left[\frac{ 1+\epsilon  (4+2\log(\pi e^{\gamma_E}) )}{2^7 \pi ^6 \epsilon ^2}+\CO(1)\right]\frac{N(N+2)\lambda_0 h_0^2}{3\eta^{2(1-2\epsilon)}}\,.
\end{align}
For the fourth diagram in Figure \ref{4diabeta} we first integrate over the two bulk points using \eqref{34MB}, and then perform the remaining integration along the defect 
\begin{align}
    \CJ_{2, 1}^{(1)} & =-\frac{N(N+2)}{3}\frac{\lambda_0^2 h_0}{\eta^{4d-14}}\frac{\pi C_{\phi}^6}{2d-7}\int \limits_{-i\infty}^{+i\infty}\frac{dz_1d z_2}{(2\pi i)^2}S\left(\frac{d-2}{2},\frac{d-2}{2},d-2; z_1,z_2\right) C_{d;\frac{d-2}{2}-z_1, \frac{d-2}{2}-z_2} \nonumber\\
    &= -\left[\frac{1+\epsilon (7/2+2\log(\pi e^{\gamma_E}) )}{2^{10} \pi ^7 \epsilon^2}+\CO(1)\right]\frac{N(N+2)\lambda^2_0 h_0}{3\eta^{2(1-2\epsilon)}}\,.
\end{align}
The last two diagrams in Figure \ref{4diabeta} admit simple analytical evaluations with the results below,
\begin{align}
\CJ_{2, 1}^{(2)}& =-\frac{N(N+2)^2}{18}\frac{\lambda_0^2 h_0}{\eta^{4d-14}}\frac{\pi C_{\phi}^6}{2d-7} C_{d;d-2, d-2}C_{d;\frac{3d}{2}-4, d-2}\nonumber\\
&= -\left[\frac{ 1+\epsilon  (3+2\log(\pi e^{\gamma_E})  )}{2^{10} \pi ^7 \epsilon ^2}+\CO(1)\right]\frac{N(N+2)^2\lambda^2_0 h_0}{6\eta^{2(1-2\epsilon)}}\,,\nonumber\\
\CJ_{2, 1}^{(3)}&=-\frac{2N(N+2)}{9}\frac{\lambda_0^2 h_0}{\eta^{4d-14}}\frac{\pi C_{\phi}^6}{2d-7} C_{d;\frac{d-2}{2},3\frac{d-2}{2}}C_{d;\frac{3d}{2}-4, \frac{d-2}{2}}   \nonumber\\
&= \left[\frac{ 1}{2^{12} \pi ^7 \epsilon }+\CO(1)\right]\frac{N(N+2)\lambda^2_0 h_0}{9\eta^{2(1-2\epsilon)}}\,.
\end{align}
Altogether, 
the one-point function, with contributions up to the third order in coupling constants of the renormalized bulk   operator $[\phi^2]_R \equiv Z_{\phi^2}^{-1} \phi_0^2$ reads
\begin{align}\label{dd}
\left\langle \left[\phi^2\right]_{R}\right\rangle_{3-\text{order}} = \frac{1}{Z_{\phi^2}}\left(A_0+A_1+B+\CJ_{0,3}+\CJ^{(1)}_{1,2}+\CJ^{(2)}_{1,2}+ \CJ_{2,1}^{(1)}+\CJ_{2,1}^{(2)}+\CJ_{2,1}^{(3)}\right)\,,
\end{align}
where  the wavefunction renormalization factor $Z_{\phi^2}$ is fully determined by the bulk theory \cite{Kleinert:2001ax}  
\begin{align}
     Z_{\phi^2} & =  1-\frac{\lambda}{(4\pi) ^2}\frac{(N+2) }{3 \epsilon }+\frac{\lambda^2(N+2)}{(4\pi) ^4}\left(-\frac{1}{3 \epsilon ^2}+\frac{5 }{36  \epsilon }\right)\nonumber\\
    & +\frac{\lambda^3(N+2)}{(4\pi) ^6}\left(-\frac{N+14}{27 \epsilon ^3}+\frac{31 N+218}{324 \epsilon ^2}-\frac{5 N+37}{108 \epsilon }\right)+\CO(\lambda^4)\,.
     \label{wavefunctionrenormorderl3}
\end{align}

\subsection{The scaling dimension of $[\hat\phi^2]_R$}
By requiring the one-point function $\left\langle \left[\phi^2\right]_{R}\right\rangle$ to be finite in the $\epsilon\to 0$ limit after making the substitution \eqref{lambdabare},
we find the  renormalization of the defect coupling to be 
\begin{gather}\label{hh0}
h_0 = \mu^\epsilon\bigg[\frac{h}{1-\frac{h}{\pi\epsilon}}+\frac{(N+2)\lambda h}{3(4\pi)^2\epsilon}-\frac{N+2}{3}\left(\frac{h^2 \lambda }{16 \pi ^3}+\frac{5 h \lambda ^2}{3\times 2^{10}\pi ^4}\right)\frac{1}{\epsilon}\nonumber\\
+\frac{N+2}{3}\left(\frac{ 3 h^2 \lambda }{2^5 \pi ^3}+\frac{(N+5)h \lambda ^2}{3\times 2^8 \pi ^4}\right)\frac{1}{\epsilon^2}\bigg]\,,
\end{gather}
which is valid to the cubic order in the couplings.
Imposing $\mu\partial_\mu h_0=0$ and using $\beta_\lambda$ from \eqref{betalambda}, we obtain the beta function of $h$,
\begin{align}
\beta_h = -\epsilon h+\frac{h^2}{\pi }+\frac{N+2}{3}\left(\frac{h \lambda }{(4 \pi) ^2}-\frac{h^2 \lambda }{(2 \pi) ^3}-\frac{5 h \lambda ^2}{6(4 \pi) ^4}\right)\,.
\end{align}
By plugging the bulk fixed point $\lambda_\star$ from \eqref{fixedlambda} into $\beta_h = 0$, we identify the defect fixed point to the $\epsilon^2$ order,
\begin{align}\label{hfixedpt}
h_\star = \frac{6 \pi  \epsilon }{N+8}+\frac{(N+2)(11N+148) \pi  \epsilon ^2}{2(N+8)^3}\,.
\end{align}
At this fixed point,  the scaling dimension of the  defect operator $[\hat\phi^2]_R$ becomes
\begin{align}\label{Dphi2}
\Delta_{\hat\phi^2} = 2+\left.\frac{\partial\beta_h}{\partial h}\right|_{h_\star, \lambda_\star} = 2+\frac{6 \epsilon }{N+8}-\frac{(N+2)(13N+44)\epsilon ^2}{2(N+8)^3}+\CO(\epsilon^3)\,,
\end{align}
whose large $N$ limit
$\Delta^{\text{Large} \,\, N} _{\hat\phi^2} = 2+\frac{6\epsilon}{N}-\frac{13\epsilon^2}{2N}+ \cdots 
$ agrees with the large $N$ analysis in \cite{Giombi:2023dqs} to  the order $\epsilon^2$,  providing a nontrivial consistency check of our result at finite $N$.

It is natural to ask how this compares with numerical results in $d=3$.
When $N=1$, setting  $\epsilon = 1$ directly in \eqref{Dphi2} truncated at the $\epsilon^2$ order gives $\Delta_{\hat\phi^2}\approx 2.549$. Instead, 
applying a resummation of the $\epsilon$-expansion in \eqref{Dphi2} using the [1,1] Pad\'e approximant yields 
\begin{align}\label{hatphi2pade11}
N=1: \quad \text{Pad\'e}_{[1,1]}\{\Delta_{\hat\phi^2}\} =\frac{2+\frac{55 \epsilon }{54}}{1+\frac{19 \epsilon }{108}}\stackrel{d=3}{=} 2.567\,. 
\end{align}
While direct numerical results for this surface critical exponent are not available, such results have been obtained for boundary critical exponents. This is where the factorization proposal of \cite{Krishnan:2023cff} (see also earlier work \cite{Bray:1977fvl}) comes in, relating the two kinds of defect critical exponents in the $d=3$ $O(N)$ CFT.

As argued in \cite{Krishnan:2023cff} based on large $N$ analysis, the surface defect with positive coupling $h_0$ flows to the ordinary fixed point, which corresponds to the
defect plane dividing the system into two disconnected regions each with a boundary $|{\rm Ord}\ra$ of the ordinary universality class \cite{Diehl:1996kd}. In other words, the conformal surface defect (interface in this case) is purely reflective and takes the factorized form $|{\rm Ord}\ra \la {\rm Ord}|$. Furthermore, 
the $O(N)$ singlet defect operator $\hat\phi^2$, which couples the two sides of the interface (irrelevant in the IR), is realized as a product of the fundamental scalar field $\hat \phi^I$ in each boundary ordinary class (see also Section~\ref{sec:fund}). 
Therefore $\Delta_{\hat\phi^2}$ is twice of $\Delta^{\rm bdry}_{\hat\phi}$. The latter is found to be 1.2751(6)  \cite{PhysRevB.83.134425} using Monte Carlo (MC) simulation and 1.23(4) \cite{Zhou:2024dbt} using fuzzy sphere regularization. Doubling the MC result gives $\Delta^{\text{MC}}_{\hat\phi^2} \approx 2.55$ for the ordinary surface defect in the $d=3$ Ising model, which agrees with [1,1] Pad\'e result \eqref{hatphi2pade11} very well. This agreement continues to hold for other small values of $N$ for which MC results are available. In Table~\ref{Dphicomp} we list our results for ${1\over 2}\Delta_{\hat\phi^2}$ from the truncated $\ep$-expansion in \eqref{Dphi2} and the [1,1] Pad\'e resummation and compare with the MC results for the boundary scaling dimension $\Delta^{\rm bdry}_{\hat\phi}$ at $N=1,2,3,4$.
This matching strongly supports the factorization  proposal of the ordinary surface defect in $d=3$.

In Appendix \ref{ordereps3}, we compute the next-order correction to the scaling dimension of $\hat\phi^2$ operator and we summarize the full result below,
\begin{align}\label{phi2hateps3order}
\left.\Delta_{\hat\phi^2} \right|_{\epsilon^3}  
=  \frac{( N+2)  \left(31552 + 6544 N + 124 N^2 + 3 N^3+96   (N+8) (5 N+22) \zeta (3) \right)}{8 (N+8)^5}\epsilon^3\,.
\end{align}
With this $\epsilon^3$ contribution, we can also apply a resummation using the [2,1] Pad\'e approximant below for $\Delta_{\hat\phi^2}$, 
\begin{gather}
    N=1: \quad \text{Pad\'e}_{[2,1]}\{\Delta_{\hat\phi^2}\} =  \frac{2+7.84282 \epsilon +2.27477 \epsilon ^2}{1+3.58808 \epsilon }\stackrel{d=3}{=}2.6411\,.
\end{gather}
This resummation shows slightly worse agreement with the Monte Carlo result for $\Delta^{\text{MC}}_{\hat\phi^2}$ compared to \eqref{hatphi2pade11} from the [1,1] Pad\'e approximant.\footnote{Performing the resummation using the Borel-Pad\'e approximant gives $\Delta_{\hat\phi^2}=2.6356$ in $3d$.} As mentioned in the introduction, a better understanding of the large order behavior of the $\epsilon$-expansion ($e.g.$ from instantons \cite{Lipatov:1976ny,Zinn-Justin:1989rgp,Dunne:2021lie,Giombi:2019upv}) will provide crucial guidance on which resummation procedure is more reliable.

\subsection{A surprising relation between $\Delta_{\phi^2}$ and $\Delta_{\hat\phi^2}$}

The scaling dimension of the bulk operator $\phi^2$, which is the lightest nontrivial $O(N)$ singlet, has been studied extensively (see \cite{Henriksson:2022rnm} for a recent summary). In particular, in the $\epsilon$-expansion, up to order $\epsilon^3$, this was computed in \cite{Kleinert:2001ax},
\ie \label{Dbulkphi211}
    \Delta_{\phi^2} &= 2 - \frac{6 \epsilon}{N+8} +\frac{(N+2) (13 N+44)\epsilon ^2}{2 (N+8)^3} \\
    &+\frac{-3 N^4+446 N^3+3576 N^2+10656 N+10624-96 (N+2) (N+8) (5 N+22) \zeta (3)}{8 (N+8)^5}\epsilon^3\,.
    \fe 
 Combining \eqref{Dphi2}, \eqref{phi2hateps3order} and \eqref{Dbulkphi211}, we arrive at the following relation,
\begin{align}\label{surp}
    \Delta_{\phi^2} + \Delta_{\hat\phi^2} = 4 +72\frac{N+2}{(N+8)^3}\epsilon^3+\CO(\epsilon^4)\,,
\end{align}
between the bulk and the defect critical exponents, in the form of a ``shadow relation'' discussed around \eqref{shadowrel} in the introduction. Note the nontrivial complete cancellations up to the $\epsilon^2$ order and the partial cancellation at the $\epsilon^3$ order for the term of leading degree in transcendentality (multiplying $\zeta(3)$).\footnote{Here we adopt the notion of degree in transcendentality (also known as the transcendental weight) defined for functions (and for special values of these functions) from iterated integrals that arise in the evaluation of massless Feynman diagrams (see $e.g.$ \cite{Brown:2009rc}). For example, the polylogarithm function ${\rm Li}_k(z)$ and the Riemann zeta value $\zeta(k)={\rm Li}_k(1)$ are both of transcendental weight $k$ in this definition. Evidence for and applications of interesting transcendentality properties in the $\epsilon$-expansion of CFT data in the $O(N)$ model were identified in \cite{Alday:2017zzv,Henriksson:2018myn,Guha:2019ipe,Ferrero:2019luz}. It would be interesting to investigate if similar transcendentality properties play any role in the non-renormalization properties we observe here for the ordinary surface defect and possible generalizations thereof.
} In Appendix \ref{checkSurp} we provide another consistency check for this relation.

The relation \eqref{surp} at $\epsilon=0$ is easily understood due to the marginally irrelevant nature of the surface coupling $h_0$ in \eqref{SDdef} of the free theory in $d=4$ (see $e.g.$ \cite{Lauria:2020emq}).
The cancellation of the order $\epsilon$ terms in \eqref{surp} can be deduced from the standard conformal perturbation theory starting from the trivial (transparent) surface defect. In fact, it is no different from the same argument for a perturbative RG flow in the bulk which we review now.
Namely, turning on a slightly relevant operator $O$ of dimension $D-\delta$ ($\delta\ll 1$) in a $D$-dimensional UV CFT triggers a short RG flow to the IR. The beta function for the renormalized coupling $g$ of this operator takes the following generic form ($e.g.$ in the absence of symmetries acting on $O$) \cite{Zamolodchikov:1986gt,Cardy:1988cwa,Klebanov:2011gs,Fei:2015oha}
\begin{align}
    \beta_g = -\delta g+ \mathcal C g^2+\CO\left(g^3\right)\,,
\end{align}
where $\mathcal C$ is a positive order 1 constant determined by the ratio of the three- and two-point functions of $O$. At the IR fixed point $g_\star \approx\frac{\delta}{\mathcal C}$. the scaling dimension of $O$ becomes
\begin{align}
    \Delta_{\rm IR} = D+\left.\partial_g \beta_g \right|_{g_\star} = D+\delta+\CO\left(\delta^2\right).
\end{align}
In our case, $D=2$ and the UV operator is  the $O(N)$ singlet $\phi^2$ at the WF fixed point (on the trivial surface), while the IR operator is $\hat\phi^2$ at the fixed point $h_\star$ describing the nontrivial ordinary surface defect. The cancellation of the $\epsilon^2$ terms in \eqref{surp} is beyond the prediction of the leading order conformal perturbation theory.

In the large $N$ limit, $\Delta_{\phi^2} + \Delta_{\hat\phi^2} = 4 $ is expected to hold to all orders in the $\epsilon$-expansion including the first $1/N$ correction \cite{Giombi:2023dqs}. The reasoning is parallel to the conformal perturbation theory argument presented above, with $1/N$ playing the role of $\delta$. The order $\epsilon^3$ term in \eqref{surp} is subleading in the $1/N$ expansion and thus does not violate this large $N$ result.

In strictly $d=3$, the surface (interface) is expected to factorize into a pair of ordinary boundary conditions.
There is a correction to this relation at the $1/N$ order, because $
    \Delta_{\phi^2} = 2-\frac{32}{3\pi^2 N}+\CO(N^{-2})$ \cite{ma19761,Lang:1992pp,Lang:1994tu,Derkachov:1997qv} and $ \Delta_{\hat\phi^2} = 2  \Delta_{\hat\phi} = 2+\frac{4}{3N}+\CO(N^{-2})$ \cite{Ohno:1983lma}, which together give\footnote{In the free $d=3$ $O(N)$ CFT, this shadow relation holds exactly without $1/N$ corrections. In this case, the surface defect factorizes into a pair of Dirichlet boundary conditions $\phi^I|_{\rm bdry}=0$ and $\hat \phi^I \propto \partial_\eta\phi^I|_{\rm bdry}$  \cite{Liendo:2012hy} and thus $\Delta_{\hat\phi^2}=2\Delta_{\hat \phi}^{\rm bdry}=3$.}
\ie 
\Delta_{\phi^2} + \Delta_{\hat\phi^2} \stackrel{d=3}{=} 4 + \frac{0.253...}{N}+\CO(N^{-2})\,.
\fe

However, we find that the approximate relation $\Delta_{\phi^2} + \Delta_{\hat\phi^2} \approx 4 $ in $d=3$ holds at much higher precision numerically even for small $N$.  For example, when $N=1$, the Monte Carlo result $\Delta^{\rm MC}_{\hat\phi^2}\approx 2.55$ (from $2\Delta^{\rm bdry}_{\hat\phi}$ \cite{PhysRevB.83.134425}) and the conformal bootstrap result $\Delta_{\phi^2}\approx 1.41$ \cite{Kos:2016ysd} add up to about 3.96. For other small values of $N$, we list the relevant numerical data of $\Delta_{\hat\phi^2}$ and $\Delta_{\phi^2}$ in Table~\ref{Dphicomp}, and we find the sum $\Delta_{\phi^2} + \Delta_{\hat\phi^2}$ to be around $3.97$ when $N=2$, $3.98$ when $N=3$ and $3.98$ when $N=4$, providing further evidence for this (approximate) non-renormalization property.

Let us also comment on a similar approximate relation for anomalous dimensions that arise for
the localized magnetic line defect (also known as the pinning field defect), defined by adding the perturbation $h_0\int_\gamma \phi_I$ along a line  $\gamma \subset \mathbb R^d$ to the bulk action of the $O(N)$ model \cite{Cuomo:2021kfm}. The conformal perturbation theory argument above (with $D=1$) does not directly apply because the three-point function of $\phi_I$ vanishes as a consequence of the $O(N)$ symmetry. Consequently the fixed point value of the renormalized coupling $h_\star$ is of order 1 instead of order $\epsilon$ \cite{Cuomo:2021kfm}. Due to the breakdown of the leading order perturbation theory, we do not expect the relation $\Delta_{\phi}+\Delta_{\hat\phi} \approx 2 $ even in the $\epsilon$-expansion (where $\Delta_{\hat\phi}$ denotes the IR scaling dimension of the operator $\hat\phi_I$ on the line). Indeed, at order $\epsilon$, we have $\Delta_\phi = 1-\frac{\epsilon}{2}$ and $\Delta_{\hat\phi} = 1+\epsilon$ \cite{Cuomo:2021kfm}.  Interestingly, if we consider directly $d=3$, the conformal bootstrap result $\Delta_\phi\approx 0.5181489(10)$ \cite{Kos:2016ysd} and the MC result $\Delta^{\rm MC}_{\hat\phi} \approx 1.52(6)$ \cite{ParisenToldin:2016szc} seem to suggest the approximate relation $\Delta_{\phi}+\Delta_{\hat\phi} \approx 2 $ in the Ising model. Together with the large $N$ result in \cite{Cuomo:2021kfm}, it appears that this approximate relation could potentially hold for all $N$.

\subsection{The defect OPE coefficient of $[\phi^2]_R \sim \mathbf{I}$}
Knowing the fixed point value of $h$ to the $\epsilon^2$ order also allows us to determine the coefficient of the one-point function $\langle [\phi^2]_R\rangle$ of the renormalized $O(N)$ singlet operator to higher order compared to  \cite{Giombi:2023dqs}. At the fixed point $(\lambda_\star, h_\star)$, this one-point function is fixed by conformal symmetry up to an overall normalization constant (denoted by $a_{\phi^2}$)
\begin{align}\label{dd1}
\left\langle  [\phi^2]_R(0,\eta)\right\rangle = \frac{a_{\phi^2}}{|\eta|^{d-2+\gamma_{\phi^2}}}\,,
\end{align}
where the anomalous dimension $\gamma_{\phi^2}$
is given by \eqref{gammaphi2}. The constant $a_{\phi^2}$ quantifies how the defect identity operator \textbf{I} contributes to the bulk-defect OPE of $[\phi^2]_R$.
To obtain $a_{\phi^2}$ to the $\epsilon^2$ order, we keep only the first three terms in \eqref{dd} and evaluate them at the fixed point,
\begin{align}\label{dd2}
\left\langle  [\phi^2]_R(0,\eta)\right\rangle
&=-\frac{3 N \epsilon}{4(N+8)\pi^2 \eta^{d-2+\gamma_{\phi^2}}}\\
&+\frac{N((N+2)(N-52)-36(N+8)\log(\pi e^{\gamma_E}))\epsilon^2}{16(N+8)^3\pi^2\eta^{d-2+\gamma_{\phi^2}}}+\CO(\epsilon^3)\nonumber \,.
\end{align}
Comparing \eqref{dd1} and \eqref{dd2} yields 
\begin{align}\label{ta2}
a_{\phi^2} = -\frac{3 N \epsilon}{4(N+8)\pi^2}+\frac{N[(N+2)(N-52)-36(N+8)\log(\pi e^{\gamma_E})]\epsilon^2}{16(N+8)^3\pi^2}+\CO(\epsilon^3)\,.
\end{align}
To compare with literature, we normalize $[\phi^2]_R$ as follows 
\begin{align}\label{CO2}
    \CO_2\equiv -\CN_{\phi^2}^{-1}[\phi^2]_R\,,
\end{align}
where $\CN_{\phi^2}$ is the square root of the two-point function coefficient of $[\phi^2]_R$ in the absence of the defect and its explicit expression is given by \eqref{phi2normalization}. The coefficient of $\langle \CO_2\rangle$, denoted by $a_{\CO_2}$, is thus
\begin{align}\label{a2}
a_{\CO_2} = -\frac{a_{\phi^2}}{\CN_{\phi^2}} = \frac{3 \sqrt{N} \epsilon }{\sqrt{2} (N+8)}+\frac{5\sqrt{N}(N+2)(N+20) \epsilon ^2}{4\sqrt{2}
   (N+8)^3}+\CO(\epsilon^3)\,.
\end{align}
Its leading  large $N$ behavior $
a_{\CO_2} =\frac{1}{\sqrt{2N}}\left(3\epsilon+\frac{5}{4}\epsilon^2\right)+\cdots 
$
agrees with  \cite{Giombi:2023dqs} up to an overall minus sign due to our convention. When $N=\epsilon=1$, the numerical value of $a_{\CO_2}$ given by the truncated $\epsilon$-expansion in \eqref{a2} is about $0.312$. Since the ordinary surface defect is a product of two decoupled ordinary boundary conditions at long distance, the surface one-point function coefficient $a_{\CO_2}$ should be the same as its boundary counterpart. For the ordinary boundary of $d=3$ Ising model, this coefficient has  recently been calculated using conformal bootstrap \cite{Gliozzi:2015qsa} and fuzzy sphere regularization \cite{Zhou:2024dbt}. For example, the fuzzy sphere result from \cite{Zhou:2024dbt} is 
\begin{align}\label{FZa}
\text{Fuzzy sphere}: \quad \langle \cO_2(0,\eta)\rangle = \frac{\tilde a_{\cO_2}}{(2 \eta)^{d-2+\gamma_{\phi^2}}}, \quad \tilde a_{\cO_2} \approx 0.74(4)\,.
\end{align}
Comparing \eqref{dd1} with \eqref{FZa}, we see the one-point function coefficients in $d=3$ are related by
\ie 
\epsilon\text{-expansion}:\quad \tilde a_{\cO_2}=2^{1+\C_{\phi^2}}a_{\cO_2}  {\approx}  \left.0.312\times 2^{2-\frac{6}{N+8}\epsilon}\right|_{N=\epsilon=1}\approx 0.786\,,
\fe
which agrees with the fuzzy sphere result within the reported uncertainty.

In Appendix~\ref{ordereps3}, we also obtain the order $\epsilon^3$ correction to $a_{\CO_2}$ in \eqref{a2eps3}. This allows us to apply the [2,1] Pad\'e resummation to $a_{\CO_2}$. For example, when $N=1$, the Pad\'e resummed result is
\begin{align}
    N=1: \quad \text{Pad\'e}_{[2,1]}\{a_{\CO_2}\} = \frac{0.235702 \epsilon+0.238687 \epsilon ^2 }{1+0.688589 \epsilon }\,,
\end{align}
which yields $a_{\CO_2}\approx 0.281$ for $\epsilon=1$. In the convention of \cite{Zhou:2024dbt}, it then gives
\ie 
\epsilon\text{-expansion~and~Pad\'e}:\quad \tilde a_{\CO_2}\approx 0.281\times 2^{\Delta_{\phi^2}}\approx 0.759\,.
\fe 
Here we have used $\Delta_{\phi^2} =2-\frac{2 \epsilon }{3} +\frac{19 \epsilon ^2}{162}$ from \eqref{Dbulkphi211} and its resummation Pad\'e$_{[1,1]}\{\Delta_{\phi^2}\}\approx 1.433$. This gives even better agreement with the fuzzy sphere result \eqref{FZa}.

\section{Fundamental Scalar in the Bulk and on the Defect}
\label{sec:fund}

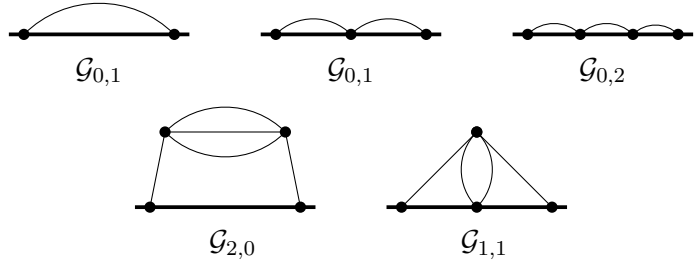
\begin{figure}[!htb]
\centering
\begin{tikzpicture} 
\draw [line width = 0.5mm] (-1.2, -1) to (1.2,-1);
\draw  (-1,-1)node[vertex]{}  to [out=45,in=135]   (1,-1)node[vertex]{};
\node at (0,-1.5) {$\CG_{0}$};
\end{tikzpicture}
\qquad 
\begin{tikzpicture} 
\draw [line width = 0.5mm] (-1.2, -1) to (1.2,-1);
\draw  (-1,-1)node[vertex]{}  to [out=45,in=135]  (0,-1)node[vertex]{}  to [out=45,in=135]  (1,-1)node[vertex]{};
\node at (0,-1.5) {$\CG_{0,1}$};
\end{tikzpicture}
\qquad 
\begin{tikzpicture} 
\draw [line width = 0.5mm] (-1.2, -1) to (1.2,-1);
\draw  (-1,-1)node[vertex]{}  to [out=45,in=135]  (-0.3,-1)node[vertex]{}  to [out=45,in=135]  (0.4,-1)node[vertex]{} to [out=45,in=135]  (1,-1)node[vertex]{};
\node at (0,-1.5) {$\CG_{0,2}$};
\end{tikzpicture}\\
\begin{tikzpicture} [baseline=-15pt]
\draw [line width = 0.5mm] (-1.2, -1) to (1.2,-1);
\draw  (-0.8,0) node[vertex]{}  to [out=45,in=135]   (0.8,0)node[vertex]{} to (-0.8, 0) to [out=-45,in=-135]   (0.8,0) ;
\draw  (-1,-1) node[vertex]{}  to (-0.8, 0);
\draw  (1,-1) node[vertex]{}  to (0.8, 0);
\node at (0.1,-1.5) {$\CG_{2,0}$};
\end{tikzpicture}
\qquad
\begin{tikzpicture} [baseline=-15pt]
\draw [line width = 0.5mm] (-1.2, -1) to (1.2,-1);
\draw  (0,-1) node[vertex]{}  to [out=135,in=-135]   (0,0)node[vertex]{}  to [out=-45,in=45]   (0,-1) ;
\draw  (-1,-1) node[vertex]{}  to (0, 0) node[vertex]{}  to  (1,- 1) node[vertex]{} ;
\node at (0.1,-1.5) {$\CG_{1,1}$};
\end{tikzpicture}
\caption{Diagrams for $\langle \hat\phi^I_0(x_1)\hat\phi^I_0(x_2)\rangle$ up to the quadratic order in the couplings. }
\label{diagrams2ptphihat}
\end{figure}

In this section, we shift our focus to the fundamental scalar $\phi^I$, which is the lightest nontrivial operator in the $O(N)$ WF CFT. The OPE of this $O(N)$ vector operator with the ordinary surface defect is dominated to leading order by the the defect fundamental field $\hat\phi^I$.
We will derive the  anomalous dimension of this defect operator to the $\epsilon^2$ order and then compute the bulk-defect two-point function $\la \phi^I\hat \phi^I\ra$. For this purpose, we need to compute the two-point function of $\hat\phi^I$ to the quadratic order in the couplings. The relevant diagrams for the bare two-point function $\langle \hat\phi^I_0(x_1)\hat\phi^I_0(x_2)\rangle$, where there is no summation over index $I$, are shown in Figure~\ref{diagrams2ptphihat}.
The first diagram is just the free propagator on the surface defect, i.e. $\CG_0 = G_d(x_1, x_2)$. The next three diagrams follow straightforwardly from the integration identity \eqref{Cdef},
\begin{align}
&\CG_{0,1}
=  -2 h_0C_\phi  C_{2;\frac{d-2}{2}, \frac{d-2}{2}} x_{12}^\epsilon G_d(x_1, x_2)\,,\nonumber\\
&\CG_{0,2}
 = 4 h_0^2C_\phi^2  C_{2;\frac{d-2}{2}, \frac{d-2}{2}}C_{2;\frac{d-2}{2}, d-3} x_{12}^{2\epsilon} G_d(x_1, x_2)\,,\\
&\CG_{2,0}
 = \frac{N+2}{3}\frac{\lambda_0^2 }{6}C_\phi^4 C_{d;\frac{d-2}{2}, 3\frac{d-2}{2}}C_{d;\frac{d-2}{2},\frac{3d}{2}-4}x_{12}^{2\epsilon} G_d(x_1, x_2)\,.\nonumber
\end{align}
The integral corresponding to the last diagram in Figure~\ref{diagrams2ptphihat} reads
\begin{align}
\CG_{1, 1}
=\frac{N+2}{3}\lambda_0 h_0 C_\phi^4\int \frac{d^2 x_0 d^2 x_3 d^{d-2}\eta}{(x_{13}^2+\eta^2)^{\frac{d-2}{2}}(x_{23}^2+\eta^2)^{\frac{d-2}{2}}(x_{03}^2+\eta^2)^{d-2}}\,.
 \end{align}
Integrating out the defect point $x_0$ gives $\frac{\pi}{(d-3)\eta^{2(d-3)}}$ and the remaining integral over the bulk point is solved in Appendix~\ref{AI},
\begin{align}
\CG_{1,1} = \frac{N+2}{3} \frac{\pi C_\phi^3}{d-3}\lambda_0 h_0 \CB\left(\frac{d-2}{2}, \frac{d-2}{2}, d-3\right)x_{12}^{2\epsilon}G_d(x_1, x_2)\,,
\end{align}
where the function $\CB$ is defined in \eqref{CB}. We define the renormalized defect operator $\hat\phi^I = Z_{\hat\phi}^{-1}\hat\phi^I_0$. Then its two-point function to this order is 
\begin{align}
\label{phihatphihat2pf}
\left\langle\hat\phi^I(x_1)\hat\phi^I(x_2)\right\rangle=\frac{G_d+\CG_{0,1}+\CG_{0,2}+\CG_{2, 0}+\CG_{1,1}}{Z_{\hat\phi}^2}\,. 
\end{align}

\subsection{The scaling dimension of $\hat\phi^I$ }
By requiring the two-point function of $\hat\phi^I$ to be finite in the limit $\epsilon\to 0$,  we obtain\footnote{The $h$ and $h\lambda$ terms  in \eqref{Zhphi} were also found in \cite{Giombi:2023dqs}.}
\begin{align}\label{Zhphi}
Z_{\hat\phi} = 1-\frac{h}{\pi\epsilon} - \frac{(N+2)\lambda^2}{72(4\pi)^4\epsilon}+\frac{2(N+2) h\lambda}{3(4\pi)^3}\left(-\frac{1}{\epsilon^2}+\frac{1}{\epsilon}\right)\,,
\end{align}
which leads to the anomalous dimension of $\hat\phi^I$ to this order,
\begin{align}
\gamma_{\hat\phi} = \beta_\lambda \partial_\lambda \log Z_{\hat\phi} + \beta_h \partial_h \log Z_{\hat\phi} = \frac{h}{\pi }+\frac{N+2}{3}\frac{\lambda (\lambda -192 \pi  h)}{3072 \pi ^4}\,.
\end{align}
At the IR fixed point, it gives
\begin{align}\label{Dphi}
\Delta_{\hat\phi} = \frac{d-2}{2} + \left.\gamma_{\hat\phi}\right|_{\lambda_\star, h_\star} = 1+\left(\frac{6}{N+8}-\frac{1}{2}\right)\epsilon+\frac{(112-N)(N+2) \epsilon^2}{4(N+8)^3}+\cO(\epsilon^3)\,.
\end{align}
In the large $N$ limit, \eqref{Dphi} becomes
\begin{align}
\Delta^{\text{Large} \,\, N} _{\hat\phi} = 1-\frac{\epsilon}{2}+\frac{6\epsilon}{N}-\frac{\epsilon^2}{4N}+\cdots\,,
\end{align}
which agrees with the large $N$ analysis in \cite{Giombi:2023dqs} to the $\epsilon^2$ order. 

To compare with numerical results in $d=3$, we apply the [1,1] Pad\'e approximant to $\Delta_{\hat\phi}$ in \eqref{Dphi} yields the resummed result below,
\begin{align}
    \text{Pad\'e}_{[1,1]}\{\Delta_{\hat\phi}\} = \frac{1+\frac{\left(-N^3-N^2+158 N+96\right) }{2 (N-4) (N+8)^2} \epsilon}{1+\frac{(N+2)(112-N)
    }{2 (N-4) (N+8)^2}\epsilon}\,.
\end{align}
There is a pole in the denominator when $N\le 3$ or $N\ge 113$. It indicates this Pad\'e approximant is not reliable for extrapolation to $d=3$ in cases of small $N$, namely the Ising model ($N=1$), XY model ($N=2$) and Heisenberg model ($N=3$). 
When $N=4$, the order $\epsilon$ term in $\Delta_{\hat\phi}$ vanishes, and hence we cannot use the [1,1] Pad\'e procedure either. Nonetheless, by directly using the truncated $\epsilon$-expansion in \eqref{Dphi} and assuming the factorization proposal for the surface defect in $d=3$,
we can compare our results with Monte Carlo simulations in the literature for the cases of $N=1,2, 3, 4$, as summarized in Table~\ref{Dphicomp}.\footnote{For $N=1$, $\Delta_{\hat\phi}$ was also computed by the fuzzy sphere regularization \cite{Zhou:2024dbt}. The result $\Delta_{\hat\phi}=1.23(4)$ is consistent with MC data within its error bar. For $N=2,3,4$, earlier MC simulation can be found in $e.g.$ \cite{Deng:2005dh, PhysRevLett.127.120603,PhysRevLett.126.135701,PhysRevE.73.056116}.} Note that the factorization of the ordinary surface defect in $d=3$ relates the various surface and the boundary scaling dimensions by 
\ie 
\Delta_{\hat\phi}={1\over 2}\Delta_{\hat\phi^2}=\Delta_{\hat\phi}^{\rm bdry}\,.
\label{factorrelation}
\fe 
From Table~\ref{Dphicomp}, we see that our results for ${1\over 2}\Delta_{\hat \phi^2}$ matches better with the Monte Carlo answers, compared to what we have for $\Delta_{\hat \phi}$, despite the exact relations in \eqref{factorrelation} from factorization. It would be interesting to understand why this is the case.

\begin{table}[!htb]
\centering
\begin{tabular}{ |c|c|c| c| c| c|} 
 \hline
$N$ &$\Delta_{\hat\phi}^{\rm bdry}$: Monte Carlo & $\Delta_{\hat\phi}$: \eqref{Dphi} & $\frac{1}{2}\Delta_{\hat\phi^2}: \eqref{Dphi2}$ & $\frac{1}{2}\Delta_{\hat\phi^2}: \text{Pad\'e}_{[1,1]}$ & $\Delta_{\phi^2}$: bootstrap\\ 
 \hline
1 & 1.2751 (6) \cite{PhysRevB.83.134425}  & 1.281 & 1.275  & 1.28346 & 1.412625(10)\cite{Kos:2016ysd} \\ 
  \hline
2 & 1.2286(25) \cite{Toldin:2023fny} & 1.21 & 1.23  & 1.24324 & 1.51136(22)\cite{Chester:2019ifh} \\ 
 \hline
 3 & 1.194(3) \cite{Toldin:2023fny} & 1.147 & 1.195  & 1.2121&1.59489(59)\cite{Chester:2020iyt}\\ 
\hline
 4 & 1.158(3) \cite{Toldin:2023fny} & 1.094 & 1.167 & 1.1875 &  1.6674(87)\cite{Kos:2013tga} \\ 
\hline
\end{tabular}
\caption{The scaling dimensions of various defect and bulk operators at the surface ordinary fixed point of the $d=3$ $O(N)$ CFT. The second column lists the Monte Carlo results in the literature. The third column is obtained by setting $\epsilon=1$ in the truncated $\ep$-expansion from \eqref{Dphi}. The next two columns show half of the scaling dimensions of $\hat\phi^2$ in $d=3$. In particular, the [1,1] Pad\'e approximant is used to compute the resummed answer in the fifth column. The last column lists the bootstrap data of the scaling dimensions of the bulk $\phi^2$ operator.}
\label{Dphicomp}
\end{table}

Finally, we determine the precise defect two-point function of the fundamental field to the $\epsilon^2$ order from \eqref{phihatphihat2pf},
\begin{gather}
    \left\langle \hat\phi^I(x_1,0)\hat\phi^I(x_2,0)\right\rangle=\frac{\CN^2_{\hat\phi}}{(x^2_{12})^{ \Delta_{\hat\phi}}}\,, 
\end{gather}
where the normalization factor is 
\begin{align}
\label{Normphihatsquare}
    \CN^2_{\hat\phi}&=\frac{1}{4 \pi ^2}+\frac{\epsilon  (N-4)\log (\pi e^{\gamma_E} )}{8 \pi ^2 (N+8)}-\frac{\epsilon ^2}{192 \pi ^2 (N+8)^3}\bigg(-12 (N-112) (N+2) \log (\pi e^{\gamma_E} ) \nonumber\\
    &+(N+8) \left(-249 (N+2)-\pi ^2 (N-4) (N+8)-6  (N-4)^2 \log ^2(\pi e^{\gamma_E} )\right)\bigg)\,.
\end{align}

\subsection{The $\langle\phi^I\hat\phi^I\rangle$ two-point function}
\label{sec:phiphih2pf}

We now determine the coefficient of $\hat\phi^I$  in the defect OPE of the bulk  field $\phi^I$. This amounts to computing the two-point function $\langle \phi^I (x_1, \eta_1)\hat\phi^I(0)\rangle$. To the $\epsilon^2$ order, the relevant diagrams are shown in Figure \ref{diagramsOPEphiphihat}.

\begin{figure}[!htb]
\centering
\begin{tikzpicture} 
\draw [line width = 0.5mm] (-1, -1) to (1,-1);
\draw   (0,0.5)node[vertex]{} to  (0,-1)node[vertex]{};
\node at (0.1,-1.5) {$\CQ_0$};
\end{tikzpicture}
\qquad 
\begin{tikzpicture} 
\draw [line width = 0.5mm] (-1, -1) to (1,-1);
\draw   (0,0.5) node[vertex]{} to (-0.5, -1)node[vertex]{} to[out=45,in=135] (0.5, -1)node[vertex]{};
\node at (0.1,-1.5) {$\CQ_1$};
\end{tikzpicture}
\qquad 
\begin{tikzpicture} 
\draw [line width = 0.5mm] (-1, -1) to (1,-1);
\draw   (0,0.5) node[vertex]{} to (-0.5, -1)node[vertex]{} to[out=45,in=135] (0, -1)node[vertex]{} to[out=45,in=135] (0.5, -1)node[vertex]{};
\node at (0.1,-1.5) {$\CQ_2$};
\end{tikzpicture}
\qquad
\begin{tikzpicture} 
\draw [line width = 0.5mm] (-1, -1) to (1,-1);
\draw   (0,0.5) node[vertex]{} to (0, -0.2)node[vertex]{} to[out=-100,in=30] (-0.5, -1)node[vertex]{} to[out=80,in=-150] (0, -0.2);
\draw (0, -0.2) to  (0.5, -1)node[vertex]{};
\node at (0.1,-1.5) {$\CQ_{1,1}$};
\end{tikzpicture}
\qquad
\begin{tikzpicture} 
\draw [line width = 0.5mm] (-1, -1) to (1,-1);
\draw   (0,0.5)node[vertex]{} to  (0,0)node[vertex]{} to (0,-0.5)node[vertex]{} to (0,-1)node[vertex]{};
\draw   (0,0)   to[out=-45,in=45] (0,-0.5) to[out=135,in=-135] (0,0);
\node at (0.1,-1.5) {$\CQ_{0,2}$};
\end{tikzpicture}
\caption{Diagrams for $\langle \phi^I \hat\phi^I\rangle$ up to the quadratic order in the couplings.}
\label{diagramsOPEphiphihat}
\end{figure}

\begin{figure}
    \centering
  \scalebox{0.65}{ \begin{tikzpicture}
 \draw[  thick] (-6,0)--(5,0);
\draw [-{Stealth[round, length=6pt, width=6pt, bend]}](-6,0) -- (5,0) node[right] {$\text{Re}(z)$};
  \draw[  thick] (0,-3)--(0,5);
\draw [-{Stealth[round, length=6pt, width=6pt, bend]}](0,-3) -- (0,5) node[above] {$\text{Im}(z)$};

    \foreach \x in {-4.5,-2} {
        \draw[thick] (\x-0.1,0.1) -- (\x+0.1,-0.1);
        \draw[thick] (\x-0.1,-0.1) -- (\x+0.1,0.1);
    }
    \filldraw[black] (0,0) circle (2pt);
    \filldraw[black] (2,0) circle (2pt);
    \filldraw[black] (4.3,0) circle (2pt);
     \node[below] at (-2.1,-0.1) {$-\frac{\epsilon}{2}$};
     \node[left] at (0,-0.4) {$0$};
     \node[below] at (2,-0.1) {$\frac{\epsilon}{2}$};
      \node[below] at (4.3,-0.1) {$1$};
       \node[below] at (-4.5,-0.1) {$-1+\epsilon$};

    \draw[thick,blue] (-0.7,-3) -- (-0.7,4.8);
    \node[below right, blue] at (-0.7,2.8) {$C_0$};
     \draw [-{Stealth[round, length=6pt, width=6pt, bend]},blue](-0.7,2.4) -- (-0.7,2.7);
    
    \draw[thick,red,] (-3,-3) -- (-3,5);
    \node[below right, red] at (-3,2.8) {$C_1$};
     \draw [-{Stealth[round, length=6pt, width=6pt, bend]},red](-3,2.4) -- (-3,2.7);
     \filldraw[red] (-3,0) circle (2pt);
        \node[left] at (-3,-0.4) {$-\frac{1}{2}$};
        
         \draw[thick,red] (-2,0) circle(0.65);
   
     \draw [-{Stealth[round, length=6pt, width=6pt, bend]},red](-1.35,-0.1) -- (-1.35,0.1);
     \node[below right, red] at (-1.6,0.9) {};
\end{tikzpicture}
}
    \caption{Contour deformation for the MB integral \eqref{MBfintshift}. The points denoted by crosses are the poles of the integrand. The poles with a $\Gamma \left(z+\cdots\right)$ dependence are denoted as crosses, while the poles with a $\Gamma\left(-z+\cdots\right)$ dependence are denoted as black nodes.}
    \label{contourdeformC0toC1}
\end{figure}
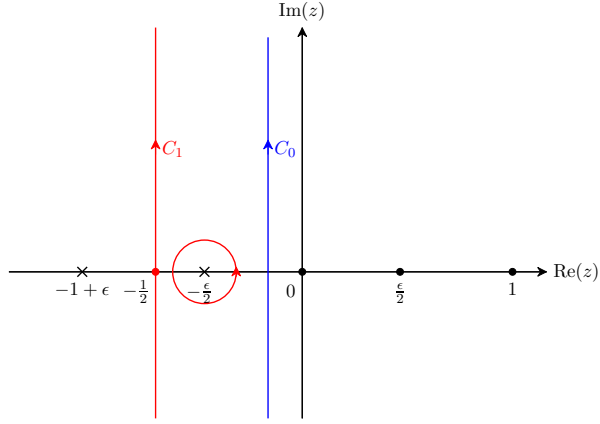

The first diagram simply evaluates to $\CQ_0  = \frac{C_\phi}{(x_1^2+\eta^2_1)^{\frac{d-2}{2}}}$. The contributions from the second and the third diagrams take a similar form 
\begin{align}
\CQ_1 &= -2 h_0 C_\phi^2\int \frac{d^2x_2 }{(x_{12}^2+\eta^2_1)^{\frac{d-2}{2}}x_2^{d-2}}\,,\nonumber\\
\CQ_2 &= 4h_0^2 C_\phi^3  C_{2;\frac{d-2}{2},\frac{d-2}{2} }\int \frac{d^2x_2 }{(x_{12}^2+\eta^2_1)^{\frac{d-2}{2}}x_2^{2(d-3)}}\,.
\end{align}
For our purpose, we need to expand $\CQ_1$ to order $\epsilon$ and $\CQ_2$ to order $\epsilon^0$. Let us carry out the calculation of $\CQ_1$ explicitly since the strategy will be useful for evaluating other diagrams. The first step is to implement Feynman parametrization, which allows us to integrate out $x_2$
\begin{align}
\CQ_1
&= -\frac{2\pi h_0 \Gamma(d-3)C_\phi^2}{\Gamma(\frac{d-2}{2})^2}\int_0^1 du \frac{u^{1-\frac{d}{2}}(1-u)^{\frac{d-4}{2}}}{(u\eta_1^2+(1-u)(\eta_1^2+x_1^2))^{d-3}}\,.
\end{align}
The next step is to transform the Feynman parameter integral into a MB integral using \eqref{basicMB},
\begin{align}\label{CQ11}
\CQ_1
&= -\frac{2\pi h_0C_\phi^2}{\Gamma(\frac{d-2}{2})^2(\eta_1^2+x_1^2)^{d-3}}\int_{C_0}\frac{dz}{2\pi i} \Gamma(-z)\Gamma(d-3+z)\int_0^1 du u^{1-\frac{d}{2}+z}(1-u)^{1-\frac{d}{2}-z}\xi^z\nonumber\\
&=-\frac{2\pi h_0C_\phi^2}{\Gamma(\frac{d-2}{2})^2(\eta_1^2+x_1^2)^{d-3}}\int_{C_0}\frac{dz}{2\pi i} \frac{\Gamma(-z)\Gamma(1-\epsilon+z)\Gamma(\frac{\epsilon}{2}+z)\Gamma(\frac{\epsilon}{2}-z)}{\Gamma(\epsilon)}\xi^z\,,
\end{align}
where $\xi \equiv  \frac{\eta_1^2}{\eta_1^2+x_1^2}$ and the contour $C_0$ is a vertical line within the strip bounded by $\Re z=0$ and $\Re z = -\frac{\epsilon}{2}$. This choice of  the contour guarantees that the $z$ and $u$ integrals in the first line of \eqref{CQ11} are finite.
 Next, we deform the contour leftwards to, say $C_1: \Re z = -\frac{1}{2}$ (see Figure~\ref{contourdeformC0toC1}). In this process, we pick up the pole at $z = - \frac{\epsilon}{2}$,
\begin{align}\label{MBfintshift}
\int_{C_0}&\frac{dz}{2\pi i} \frac{\Gamma(-z)\Gamma(1-\epsilon+z)\Gamma(\frac{\epsilon}{2}+z)\Gamma(\frac{\epsilon}{2}-z)}{\Gamma(\epsilon)}\xi^z\\
&=\Gamma\left(\frac{\epsilon}{2}\right)\Gamma\left(1-\frac{3\epsilon}{2}\right)\xi^{-\frac{\epsilon}{2}}+\frac{1}{\Gamma(\epsilon)}\int_{C_1} \frac{dz}{2\pi i}\Gamma(-z)\Gamma(1-\epsilon+z)\Gamma\left(\frac{\epsilon}{2}+z\right)\Gamma\left(\frac{\epsilon}{2}-z\right)\xi^z\,.\nonumber
\end{align}
where  the integral along $C_1$ is finite in the $\epsilon\to0$ limit. Since $\frac{1}{\Gamma(\epsilon)}$ is order $\epsilon$, we can simply set $\epsilon=0$ in the $C_1$ integral.
Altogether, up to order $\epsilon^2$ corrections, $\CQ_1$ becomes 
\begin{align}\label{MBfintshift1}
\CQ_1 &= -\frac{2\pi h_0C_\phi^2}{(\eta_1^2+x_1^2)^{d-3}}\left[\frac{2}{\epsilon }-\log (\xi )+\epsilon  \left(\frac{\log ^2(\xi )}{4}+\frac{\pi ^2}{3}+q(\xi)\right)\right]\,, \nonumber\\ 
q(\xi) &\equiv \int_{C_1} \frac{dz}{2\pi i}\frac{\pi^2 \xi^z }{\sin^2(\pi z) z }\,.
\end{align}
\noindent{}The integral $q(\xi)$ can be computed by closing the contour on the right because $0<\xi<1$,
\begin{align}\label{hxi}
q(\xi)  =\text{Li}_2(\xi ) -\frac{1}{2} \log ^2(\xi )+\log (1-\xi ) \log (\xi )-\frac{\pi ^2}{3}\,.
\end{align}
The explicit form of $q(\xi)$ is actually not important for our purpose. The reason is that the two-point function $\langle \phi^I\hat\phi^I\rangle$  at the fixed point is completely determined by the conformal symmetry up to an overall constant. It cannot depend on complicated functions like $\text{Li}_2(\xi )$. Therefore $q(\xi)$ will eventually drop out after plugging the fixed point value into $\langle \phi^I\hat\phi^I\rangle$.

The diagram $\CQ_2$ can be evaluated using exactly the same method and we find its $\epsilon$-expansion to be 
\begin{align}
\CQ_2 = \frac{8\pi^2 h_0^2 C_\phi^3}{(x_1^2+\eta_1^2)^{\frac{3d}{2}-5}}\left(\frac{2}{\epsilon ^2}-\frac{2 \log (\xi )}{\epsilon }+\log ^2(\xi )+\pi ^2+3 q(\xi)\right)\,.
\end{align}
For the diagram $\CQ_{1,1}$, which produces the following integral, 
\begin{align}\label{sp0}
\CQ_{1,1} = \frac{(N+2)\lambda_0 h_0 C_\phi^4}{3}\int \frac{d^2 x_2 d^{d-2}\eta_2 d^2 x_3}{(x_{12}^2+\eta_{12}^2)^{\frac{d-2}{2}}(x_{23}^2+\eta_{2}^2)^{d-2}(x_{2}^2+\eta_{2}^2)^{\frac{d-2}{2}}}\,,
\end{align}
it is more involved to derive its MB representation.
First, integrating out $x_3$ gives $\frac{\pi}{d-3}\frac{1}{(\eta_2^2)^{d-3}}$. Then the $x_2$ integral can be transformed into a Feynman parameter integral
\begin{align}\label{sp2}
\int \frac{d^2 x_2 }{(x_{12}^2+\eta_{12}^2)^{\frac{d-2}{2}}(x_{2}^2+\eta_{2}^2)^{\frac{d-2}{2}}} = \frac{\pi \Gamma(d-3)}{\Gamma(\frac{d-2}{2})^2}\int_0^1 d^2 u \frac{\delta(1-u_1-u_2)(u_1 u_2)^{\frac{d-4}{2}}}{[(\eta_2-u_1 \eta_1)^2+u_1u_2(x_1^2+\eta_1^2)]^{d-3}}\,.
\end{align}
An important step is making the substitution $\eta_2\to u_1\eta_2$, after which we apply MB representation \eqref{basicMB} to the denominator of \eqref{sp2}. The MB representation allows us to compute the Feynman parameter integrals and the $\eta_2$ integral analytically. Altogether, the remaining MB integral is 
\begin{align}
\CQ_{1,1}& =\frac{(N+2)\pi\lambda_0 h_0 C_\phi^3}{3(\eta_1^2)^{\frac{d}{2}-2}(\eta_1^2+x_1^2)^{d-3}}\nonumber\\
&\times\int_{\widetilde{C}_0}\frac{dz}{2\pi i}\frac{\Gamma \left(\frac{\epsilon }{2}\right) \Gamma (z-\epsilon +1) \Gamma \left(-z-\frac{\epsilon }{2}\right) \Gamma \left(z-\frac{\epsilon }{2}+1\right) \Gamma
   \left(\frac{\epsilon }{2}-z\right) \Gamma \left(z+\frac{3 \epsilon }{2}\right)}{4\Gamma (z+1) \Gamma (2-\epsilon ) \Gamma \left(1-\frac{\epsilon
   }{2}\right) \Gamma (2 \epsilon )} \xi^z\,.
\end{align}
where the contour $\widetilde{C}_0$ is between $\Re z = -\frac{3}{2}\epsilon$ and  $\Re z = -\frac{1}{2}\epsilon$. This contour choice is required by the convergence of the $u$ and $\eta_2$ integrals. Using the same contour deformation argument as before, we obtain the $\epsilon$-expansion of $\CQ_{1,1}$,
\begin{align}
\CQ_{1,1} = \frac{(N+2)\pi\lambda_0 h_0 C_\phi^3}{3(\eta_1^2)^{\frac{d}{2}-2}(\eta_1^2+x_1^2)^{d-3}}\left(\frac{1}{2 \epsilon ^2}+\frac{2-3 \log (\xi )}{4 \epsilon }+\frac{9 \log ^2(\xi )}{16}-\frac{3 \log (\xi )}{4}+\frac{\pi ^2}{3}+\frac{1}{2}+q(\xi)\right)\,.
\end{align}
The evaluation of the last diagram $\CQ_{0, 2}$ follows from \eqref{Cdef},
\begin{align}
\CQ_{0, 2} = \frac{N+2}{3} \frac{\lambda_0^2 C_\phi^5}{6} \frac{C_{d;\frac{d-2}{2}, 3\frac{d-2}{2}}C_{d;\frac{d-2}{2}, 3\frac{d}{2}-4}}{(x_1^2+\eta_1^2)^{\frac{3}{2}d-5}}\,.
\end{align}

Summing up all five diagrams and dividing the result by the factor $\sqrt{Z_{\phi}}Z_{\hat\phi}$ (see \eqref{Zhphi} and \eqref{Zphi}), we obtain the renormalized two-point function  $\langle \phi^I(X_1)\hat\phi^I(X_2)\rangle$. It is more conventional to normalize the bulk field $\phi^I$ by $\CN_\phi$, c.f. \eqref{CNphibulk}, and the defect field $\hat\phi^I$ by $\CN_{\hat\phi}$, c.f. \eqref{Normphihatsquare}, so that their two-point functions with themselves have unit coefficients.
Altogether, after adopting this normalization, we find 
\begin{align}\label{surp2}
\frac{\langle \phi^I(x, \eta) \hat\phi^I(0)\rangle}{\CN_\phi \,\CN_{\hat\phi}} &= \left.\frac{\CQ_0+\CQ_1+\CQ_2+\CQ_{1,1}+\CQ_{0,2}}{\sqrt{Z_{\phi}}Z_{\hat\phi}\CN_\phi \CN_{\hat\phi}}\right|_{\lambda_\star, h_\star} = \frac{1+\CO(\epsilon^3)}{\eta^{\Delta_\phi-\Delta_{\hat\phi}}(x^2+\eta^2)^{\Delta_{\hat\phi}}}\,,
\end{align}
where $\Delta_\phi = 1-\frac{\epsilon}{2}+\frac{(N+2) \epsilon ^2}{4 (N+8)^2}+\CO(\epsilon^3)$, and $\Delta_{\hat\phi}$ is given by  \eqref{Dphi}. Note the surprising cancellation up to the $\ep^2$ order for this normalized two-point function. It would be interesting to see if this can be understood using other methods such as analytic bootstrap (see \cite{Bissi:2022mrs} for a recent review).

In the convention of \cite{Zhou:2024dbt} that studies the ordinary boundary for the $d=3$ Ising CFT  numerically using the fuzzy sphere method, our prediction for the corresponding boundary OPE coefficient $b^{\rm bdry}_{\phi\hat\phi}$  from the truncated $\ep$-expansion result in \eqref{surp2}  is thus 
\ie 
\epsilon\text{-expansion}:\quad  b^{\rm bdry}_{\phi\hat\phi}= \left. 2^{\Delta_\phi-\Delta_{\hat\phi}} \right|_{N=\epsilon=1} \approx0.586 \,.
\fe
 In comparison, the fuzzy sphere result is $b^{\rm bdry}_{\phi\hat\phi}\approx 0.87$ for the ordinary boundary \cite{Zhou:2024dbt}.

\section{Displacement Operators and the Zamolodchikov Norm}\label{disO}
\begin{figure}[!htb]
\centering
\begin{tikzpicture} 
\draw [line width = 0.5mm] (-1.2, -1) to (1.2,-1);
\draw    (-1,-1)node[vertex]{}  to [out=30,in=150]  (1,-1)node[vertex]{} to  [out=100,in=80] (-1,-1);
\node at (0.1,-1.5) {$I_{\rm D}^{(1)}$};
\end{tikzpicture}
\qquad
\begin{tikzpicture} 
\draw [line width = 0.5mm] (-1.2, -1) to (1.2,-1);
\draw    (-1,-1)node[vertex]{}  to [out=80,in=100]  (1,-1)node[vertex]{};
\draw    (-1,-1) to  [out=30,in=150] (0,-1)node[vertex]{} to  [out=30,in=150] (1,-1);
\node at (0.1,-1.5) {$I_{\rm D}^{(2)}$};
\end{tikzpicture}
\qquad
\begin{tikzpicture} 
\draw [line width = 0.5mm] (-1.2, -1) to (1.2,-1);
\draw    (1,-1)node[vertex]{} to  [out=100,in=80] (-1,-1) node[vertex]{} to  [out=10,in=-100] (0,-0.43)node[vertex]{} to  [out=-80,in=170] (1,-1);  
\node at (0.1,-1.5) {$I_{\rm D}^{(3)}$};
\end{tikzpicture}
\caption{Renormalization of the operator $\partial_{a} \phi_0^2$ on the defect.}
\label{DD}
\end{figure}
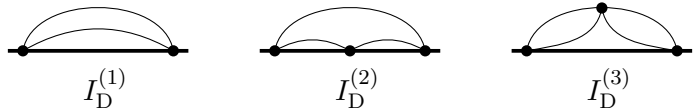

The defect breaks the translational symmetry of the bulk theory. This symmetry breaking leads to protected primary operators on the defect, known as displacement operators \cite{Billo:2016cpy}. More precisely, the displacement operator $\text{D}_a$ can be defined via the modified conservation law for the bulk stress energy tensor
\begin{align}\label{Ddef}
    \partial_\mu T^{\mu}_{\,\, a} = - \delta(\Sigma) \,\text{D}_a\,, 
\end{align}
where $\delta(\Sigma)$ denotes the delta-function supported on the defect. This definition \eqref{Ddef} has several interesting implications. First, for a $p$-dimensional defect, the displacement operator has the protected scaling dimension $p+1$. Second, given a generic bulk primary operator $O$ of dimension $\Delta_O$, \eqref{Ddef} implies the following fundamental Ward identity \cite{Billo:2016cpy},
\begin{align}\label{Ward}
\Delta_O a_O = \left(\frac{\pi}{4}\right)^{\frac{p}{2}} \frac{\sqrt{\pi}}{\Gamma(\frac{p+1}{2})} b_{O\rm D}\,,
\end{align}
relating  the one-point function coefficient $a_O$ and the bulk-defect two-point function coefficient $b_{O\rm D}$ with the displacement operator.
This is easy to verify directly in the free scalar theory \cite{Billo:2016cpy}. Third, with the normalization of ${\rm D}_a$ fixed via \eqref{Ddef} and the canonically normalized bulk stress energy tensor, the two-point function coefficient of ${\rm D}_a$ is physical and defines the Zamolodchikov norm $C_{\rm D}$ \cite{Bianchi:2015liz}, 
\begin{align}
    \langle {\rm D}_a (x_1) {\rm D}_b(x_2)\rangle = \delta_{ab}\frac{C_{\rm D}}{(x_{12}^2)^{p+1}}\,.
\end{align}
It also determines one of the defect conformal anomaly coefficients \cite{Bianchi:2015liz}, as reviewed in the introduction (see \eqref{confanom} and \eqref{d1CD}).

In this section, we will identify the displacement operator for the planar surface defect \eqref{SDdef} in the critical $O(N)$ CFT. We then compute its scaling dimension and confirm that it is protected. We also determine
the Zamolodchikov norm and test the Ward identity \eqref{Ward} for the $\phi^2$ operator in bulk.

\subsection{Identifying the displacement operator and its Zamolodchikov norm}
Using the definition \eqref{Ddef}, it is easy to see that the displacement operator in our setup is the renormalized operator ${\rm D}_a =-h\,[\partial_a \hat\phi^2]_R$, where $\hat\phi^2=\hat\phi^I \hat\phi^I$, and  $\partial_a = \partial_{\eta^a}$ is along one of the transverse directions to the surface. 
We first carry out the renormalization of this operator and derive its anomalous dimension to the leading order in $\epsilon$.

For this purpose, we compute the 
bare two-point function $\langle \partial_{a} \hat\phi_0^2(x_1)\partial_{a} \hat\phi_0^2(x_2)\rangle$ (no summation over $a$). To the first nontrivial order, there are three diagrams that contribute as shown in Figure~\ref{DD}. A short calculation gives their evaluations,
\begin{align}
I_{\rm D}^{(1)} = & \frac{4N(d-2)C_\phi^2}{x_{12}^{2(d-1)}}\,, \quad
I_{\rm D}^{(2)}=-\frac{8N(d-2)C_\phi^3  C_{2;\frac{d-2}{2}, \frac{d-2}{2}}}{x_{12}^{3(d-2)}}h_0\,,\nonumber\\
&I_{\rm D}^{(3)}=-\frac{N(N+2)}{3}\frac{(3d-8) C_\phi^4 C_{d;d-2, d-2}}{x_{12}^{3(d-2)}}\lambda_0\,.
\end{align}
We introduce the renormalization factor via
$ [\partial_{a} \hat\phi^2 ]_R =Z_{\rm D}^{-1} \partial_{a} \hat\phi_0^2$. Imposing the finiteness of the renormalized two-point function $Z_{\rm D}^{-2}(I_{\rm D}^{(1)} +I_{\rm D}^{(2)} +I_{\rm D}^{(3)})$ in the $\ep \to 0 $ limit leads to
\begin{align}
Z_{\rm D} = 1-\frac{h}{\pi \epsilon}- \frac{N+2}{3}\frac{\lambda}{(4\pi)^2 \epsilon}\,,
\end{align}
from which we can deduce the anomalous dimension of ${\rm D}_a$,
\begin{align}
\gamma_{\rm D} = \left(\beta_\lambda\partial_\lambda+\beta_h \partial_h\right) \log Z_{\rm D} = \frac{h}{\pi}+ \frac{N+2}{3}\frac{\lambda}{(4\pi)^2}\,.
\end{align}
Plugging in the bulk fixed point 
\eqref{fixedlambda}, and the defect fixed point 
\eqref{hfixedpt} while keeping only the leading order in $\epsilon$, yields 
\begin{align}
\Delta_{\rm D} = d-1+\left.\gamma_{\rm D} \right|_{\lambda_\star, h_\star} = d-1+\epsilon = 3\,.
\end{align}
Therefore, at the $\epsilon$ order, the operator $[\partial_a \hat\phi^2]_R$ indeed has the protected scaling dimension 3 as predicted by \eqref{Ddef}. To the same order, its two-point function  takes the form 
\begin{align}
\left\langle [\partial_a \hat\phi^2]_R(x_1) [\partial_b \hat\phi^2]_R(x_2)\right\rangle = \delta_{ab}\frac{\CN^2_{\partial_{\perp}\hat\phi^2}}{x^6_{12}}\,, \quad  \CN^2_{\partial_{\perp}\hat\phi^2}=  \frac{N\left(1-\frac{N+5}{N+8}\epsilon\right)}{2\pi^4 }\,.
\end{align}
For the displacement operator canonically normalized by \eqref{Ddef}, the corresponding two-point function coefficient, namely the Zamolodchikov norm, is thus
\ie 
C_{\rm D}={h_\star^2\, \CN^2_{\partial_{\perp}\hat\phi^2}}  =  \frac{18 N \epsilon ^2}{\pi ^2 (N+8)^2}+\frac{3 N \left(5 N^2+92 N+56\right) \epsilon ^3}{\pi ^2 (N+8)^4} +\cO(\ep^4)\,.
\label{CDexp}
\fe 
where we have used the fixed point value $h_\star$ in \eqref{hfixedpt}.

At $d=3$, according to the factorization proposal \cite{Krishnan:2023cff}, the surface defect factorizes into a pair of ordinary boundaries for the bulk CFT. Correspondingly, there are two protected dimension 3 defect operators, from the canonically normalized displacement operators ${\rm D}^{\rm bdry}_1$ and ${\rm D}^{\rm bdry}_2$ for each of the boundaries. The displacement operator we study here, is odd under the transverse reflection, and explicitly given by the following linear combination\footnote{It would be interesting to identify the other dimension 3 protected operator ($i.e.$  ${\rm D}'={\rm D}^{\rm bdry}_1+{\rm D}^{\rm bdry}_2$) in $d=3$ from the $\ep$-expansion. We leave this to future work.}
\ie 
{\rm D}={\rm D}^{\rm bdry}_1-{\rm D}^{\rm bdry}_2\,,
\fe
therefore, we expect the relation
\ie 
C_{\rm D}=2C_{\rm D}^{\rm bdry}\,,
\fe
between the Zamolodchikov norm for the ordinary surface versus that of the ordinary boundary in $d=3$.
For $N=1$, the fuzzy sphere result for the boundary  Zamolodchikov norm is $C_{\rm D}^{\rm bdry}=0.0089(2)$, which translates to $ C_{\rm D}\approx 0.018$.
On the other hand, our $\ep$-expansion result \eqref{CDexp}
give $C_{\rm D} \approx 0.0225$ from the $\ep^2$ term and including the $\epsilon^3$ correction yields $C_{\rm D} \approx 0.0296$. This suggests that a resummation ($e.g.$ Pad\'e after including the next order correction) is needed in order to produce a reliable prediction for $\ep=1$.

\subsection{Verifying the fundamental Ward identity}
To find the OPE coefficient between the bulk operator $O=[\phi^2]_R$ and the displacement operator, we need to compute the two-point function $\langle \phi^2_0(x_1, \eta_1)\partial_a\hat\phi^2_0(0)\rangle$. The free part of this  two-point function is 
\begin{align}
I_{O\rm D}^{(0)} = 2N C_\phi^2 \left.\partial_{\eta_2^a}\right |_{\eta_2=0}\frac{1}{(x_1^2+\eta_{12}^2)^{d-2}} = N \frac{4(d-2)C_\phi^2 \eta_1^a}{(x_1^2+\eta_1^2)^{d-1}}\,.
\end{align}
At the leading order in the bulk coupling, after integrating out the bulk vertex we have 
\begin{align}
I_{O\rm D}^{(1)} &= -\frac{N(N+2)}{3}\lambda_0C_\phi^4 C_{d;d-2, d-2}\left.\partial_{\eta_2^a}\right |_{\eta_2=0}\frac{1}{(x_1^2+\eta_{12}^2)^{\frac{3d}{2}-4}}\nonumber\\
&=-\frac{N(N+2)\lambda_0}{3}\frac{(3d-8) C_{d;d-2, d-2} C_\phi^4 \eta_1^a}{(x_1^2+\eta_1^2)^{3(\frac{d}{2}-1)}}\,.
\end{align}
The calculation of the order $h_0$ term proceeds in exactly the same way as for $\CQ_1$ in 
Section~\ref{sec:phiphih2pf},
\begin{align}
I_{O\rm D}^{(2)} &= - 8 h_0 N C_\phi^3 \left.\partial_{\eta_2^a}\right |_{\eta_2=0}\frac{1}{(x_1^2+\eta_{12}^2)^{\frac{d-2}{2}}}\int\frac{d^2 x_3}{(x_{13}^2+\eta^2_1)^{\frac{d-2}{2}}(x_3^2+\eta_2^2)^{\frac{d-2}{2}}}\nonumber\\
&=\frac{- 8(d\!-\!2) h_0 N C_\phi^3 \eta_1^a}{(x_1^2+\eta_1^2)^{\frac{d}{2}}} \int  \frac{d^2 x_3}{(x_{13}^2\!+\!\eta^2_1)^{\frac{d-2}{2}}x_3^{d-2}}\\
&=\frac{- 8\pi(d-2) h_0 N C_\phi^3 \eta_1^a}{(x_1^2+\eta_1^2)^{\frac{3}{2}(d-2)}}\left[\frac{2}{\epsilon}-\log\xi+\CO(\epsilon)\right]\,,\nonumber
\end{align}
where $\xi =\frac{\eta_1^2}{x_1^2+\eta_1^2}$.
The fully renormalized two-point function at the fixed-point is thus
\begin{gather}\label{pdp}
\left\langle [\phi^2]_R(x, \eta) [\partial_a \hat \phi^2]_R(0)\right\rangle = \frac{1}{Z_{\phi^2} Z_{\rm D}}\left.\left(I_{O\rm D}^{(0)}+I_{O\rm D}^{(1)}+I_{O\rm D}^{(2)}\right)\right|_{\lambda_\star, h_\star}\nonumber\\
=\frac{N(1-\epsilon\frac{N+5-3 \log(\pi e^{\gamma_E})}{N+8})\eta^a}{2 \pi ^4 \eta^{\Delta_{\phi^2}-2}\left(\eta ^2+x^2\right)^3}\,,
\end{gather}
where $\Delta_{\phi^2} = d-2+\gamma_{\phi^2}$ is the scaling dimension of $\phi^2$. The anomalous dimension $\gamma_{\phi^2}$ is given explicitly in \eqref{gammaphi2}. If we normalize the mass operator $\phi^2$ as in \eqref{CO2}, the above becomes,
\begin{align}\label{bphi2D}
&\left\langle \CO_2(x_1, \eta_1) {\rm D}_a(0)\right\rangle=\frac{b_{\CO_2 \rm D}\eta^a}{\eta^{\Delta_{\phi^2}-2}\left(\eta ^2+x^2\right)^3}\,,\nonumber\\
 &b_{\CO_2 \rm D}= \frac{\sqrt{N}(2-\epsilon)h_\star }{\sqrt{2}\pi^2}=\frac{6 \sqrt{2} \sqrt{N} \epsilon }{\pi  (N+8)}+\frac{\sqrt{N} \left(5 N^2+74 N-88\right) \epsilon ^2}{\sqrt{2} \pi 
   (N+8)^3}+\CO(\epsilon^3)\,.
\end{align}
Combining \eqref{a2} and \eqref{bphi2D}, we reproduce the Ward identity \eqref{Ward} to the $\epsilon^2$  order for $p=2$,
\begin{align}
\Delta_{\CO_2}a_{\CO_2} -\frac{\pi}{2}b_{\CO_2 \rm D} = \CO(\epsilon^3)~,
\end{align}
where $\Delta_{\CO_2}=\Delta_{\phi^2}$ and is given in \eqref{Dbulkphi211}.

In $d=3$, the bulk-defect two-point function coefficient $b_{\CO_2 \rm \tilde D}$ for the ordinary boundary condition of the Ising model was computed in 
\cite{Zhou:2024dbt} using fuzzy sphere regularization. There the displacement operator $\rm\tilde  D$  is normalized to have a unit two-point function coefficient and the result is \cite{Zhou:2024dbt} ,
\begin{align}
\text{Fuzzy sphere}: \quad \langle \CO_2 (0,\eta) \rm \tilde D(0)\rangle = \frac{b^{\text{bdry}}_{\CO_2 \tilde D}}{(2\eta)^{\Delta_{\phi^2}-3} \eta^6}\,, \quad  b^{\text{bdry}}_{\CO_2 \tilde D} = 0.92(4)\,,
\end{align}
  Switching to this convention, our $\ep$-expansion result \eqref{bphi2D} leads to
  \ie 
 \quad  b^{\text{bdry}}_{\CO_2  \rm  \tilde D} 
  =2^{\Delta_{\phi^2}-3}\frac{\sqrt{N}(2-\epsilon)}{\sqrt{2}\pi^2 \CN_{\partial_\perp\hat \phi^2}}  = 2^{-1-{6\ep\over N+8}} \left(2-\frac{3\epsilon}{N+8}\right)
+\CO(\epsilon^2)
  \fe 
  If we set $N=1$ and $\ep=1$ in the truncated $\ep$-expansion above, it gives 
  $b^{\text{bdry}}_{\CO_2  \rm  \tilde D}  \approx 0.525 $. On the other hand, neglecting the order $\epsilon$ correction, the leading contribution gives 1, which is closer to the fuzzy sphere result. This suggests that higher order corrections and resummation are needed to obtain a reliable prediction from the $\ep$-expansion for $\ep=1$.

\section{Defect $b$ Conformal Anomaly and the $b$-Theorem}\label{bcal}

In this section we study the $b$ conformal anomaly for the ordinary surface defect. As reviewed in the introduction, this quantity plays an important role in defect dynamics as it quantifies the defect degrees of freedom and provides a defect RG monotone.

The strategy is to analyze the free energy of the defect wrapping a sphere of radius $R$ following \cite{Giombi:2023dqs} and we will extend their results to the $\epsilon^4$ order. The $b$ anomaly controls the logarithmic divergence of the defect free energy $\cF$ defined below,
\begin{gather}\label{CFdef}
    \CF \equiv  -\log{\frac{Z_{\rm DCFT}}{Z_{\rm CFT}}}=a_1+a_2 (\mu R)^2-\frac{b}{3} \log{ (\mu R)}\,,
\end{gather}
where $Z_{\rm DCFT}$ is the partition function of the CFT in the presence of the spherical defect,
and $Z_{\rm CFT}$ is the bulk partition function. The renormalization scale is denoted by $\mu$, and the coefficients $a_1$,$a_2$ are scheme dependent ($e.g.$ in the dimensional regularization scheme the coefficient $a_2$ vanishes) whereas the coefficient $b$ is scheme independent.  This definition of $b$ is consistent with that in \eqref{confanom} from the trace anomaly due to the relation $-R\partial_R \CF = \int d^d X \,\langle T^\mu_\mu\rangle$, where the RHS is localized to an integral over the $S^2$ of radius $R$ on which the defect lives.

Up to the $\epsilon^3$ order, the results for $\CF$ derived in \cite{Giombi:2023dqs} can be summarized as 
\ie 
&\CF=-F_0-F_1-F_2\,,
\fe
with the following contributions at each order,
\ie 
&F_0 = NC_\phi^2 h_0^2 R^{2\epsilon} \CA_2(d-2)\,,\quad F_1 = - \frac{4N}{3}C_\phi^3 h_0^3 R^{3\epsilon} \CA_3\left(\frac{d-2}{2},\frac{d-2}{2},\frac{d-2}{2}\right)\,,\\
&F_2 = -\frac{N(N+2)}{6}C_\phi^4 \lambda_0 h_0^2 R^{3\epsilon}C_{d;d-2,d-2} \CA_2\left(\frac{3}{2}d-4\right)\,.
\fe 
The explicit expressions for $C_{n;a,b}$, $\CA_2$ and $\CA_3$ 
can be found in \eqref{Cdef}, \eqref{CardyA2}
and \eqref{CAfinal} respectively.

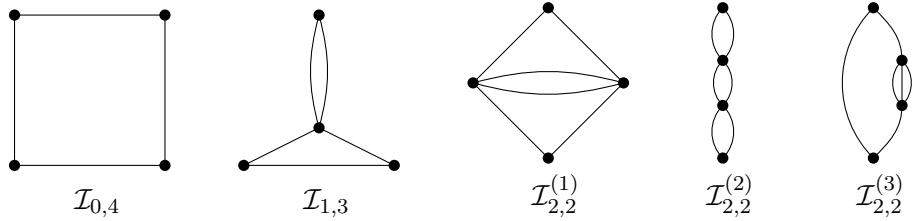
\begin{figure}[!htb]
\centering
\begin{tikzpicture} 
\draw   (-1,1)node[vertex]{} to  (1,1)node[vertex]{}  to  (1,-1)node[vertex]{}  to  (-1,-1)node[vertex]{} to  (-1,1);
\node at (0.1,-1.5) {$\CI_{0,4}$};
\end{tikzpicture}
\qquad 
\begin{tikzpicture} 
\draw   (0,1) node[vertex]{} to[out=-75,in=75]  (0,-0.5)node[vertex]{} to[out=105,in=-105]  (0,1);
\draw   (0,-0.5) to (-1,-1)node[vertex]{} to (1,-1)node[vertex]{} to (0,-0.5)  ; 
\node at (0.1,-1.5) {$\CI_{1,3}$};
\end{tikzpicture}
\qquad
\begin{tikzpicture} 
\draw   (-1,0) node[vertex]{} to[out=15,in=165]  (1,0)node[vertex]{};
\draw   (-1,0) to[out=-15,in=-165]  (1,0);
\draw   (0,1) node[vertex]{} to  (-1,0) to (0,-1) node[vertex]{} to (1,0) to (0,1);
\node at (0.1,-1.5) {$\CI_{2,2}^{(1)}$};
\end{tikzpicture}
\qquad
\begin{tikzpicture} 
\draw   (0,1) node[vertex]{} to[out=-45,in=45]  (0,0.3)node[vertex]{} to[out=-45,in=45]  (0,-0.3)node[vertex]{} to[out=-45,in=45]  (0,-1)node[vertex]{} to[out=135,in=-135]  (0,-0.3) to[out=135,in=-135]  (0,0.3) to[out=135,in=-135]  (0,1);
\node at (0.1,-1.5) {$\CI_{2,2}^{(2)}$};
\end{tikzpicture}
\qquad
\begin{tikzpicture} 
\draw   (0,1) node[vertex]{} to[out=-135,in=135]  (0,-1)node[vertex]{} to [out=45,in=-90] (0.38,-0.3) node[vertex]{} to (0.38,0.3) node[vertex]{} to [out=90,in=-45](0,1);
\node at (0.1,-1.5) {$\CI_{2,2}^{(3)}$};
\draw  (0.38,-0.3) to [out=45,in=-45] (0.38,0.3)  [out=-135,in=135] to (0.38,-0.3); 
\end{tikzpicture}
\caption{Diagrams contributing to the $b$ anomaly at the $\epsilon^4$ order.
}
\label{4dia}
\end{figure}

At the 
fourth order in couplings, the five diagrams in Figure~\ref{4dia} contribute to the $b$ anomaly.\footnote{There are also diagrams that are linear in $h$ at the same total order in the couplings. However, the same set of diagrams contribute to the one-point function of $\phi_0^2$ without a defect and hence should vanish.}  The first diagram $\CI_{0,4}$ involves four integrals on $S^2$,
\begin{align}
\CI_{0,4} = 2N h_0^4 C_\phi^4 R^{4\epsilon}\int_{S^2} \frac{d^2 X_1d^2 X_2d^2 X_3d^2 X_4}{X_{12}^{d-2}X_{23}^{d-2}X_{34}^{d-2}X_{14}^{d-2}}\,.
\end{align}
We can use the SO$(3)$ symmetry to get rid of one integral, $e.g.$ by taking $X_4 = (0,0,1,0,\cdots,0)$, which corresponds to placing the point at the origin in the stereographic coordinates. Using stereographic coordinates for the remaining coordinates, we then have
\begin{align}
\CI_{0,4} = \frac{8N\pi}{ 2^{d-2}}\pi h_0^4C_\phi^4 R^{4\epsilon}\int\prod_{a=1}^3 \left(d^2 y_a \Omega_a^2\right) \frac{\left(\Omega_1 \Omega_2 \Omega_3\right)^{2-d}}{( y_{12}^2)^{\frac{d-2}{2}}( y_{23}^2)^{\frac{d-2}{2}}(  y_{3}^2)^{\frac{d-2}{2}}( y_{1}^2)^{\frac{d-2}{2}}}\,,
\end{align}
where $\Omega_a$ is a shorthand notation for the Weyl factor $\Omega(y_a)$, which is defined in Section~\ref{nac}. Then we perform the inversion transformation $y_a^i\to y_a^i/y_a^2$. As a result of the transformation laws \eqref{inversion}, we obtain
\begin{align}\label{yyy}
\CI_{0,4} = 2N \pi h_0^4C_\phi^4 (2R)^{4\epsilon}\int \frac{d^2 y_1 d^2 y_2 d^2 y_3}{(1+y_1^2)^\epsilon (1+y_2^2)^\epsilon (1+y_3^2)^\epsilon y_{12}^{d-2}y_{23}^{d-2}}\,.
\end{align}
By applying \eqref{G1int} to the $y_3$ integral and \eqref{G0int} to the resulting $y_1$ and $y_2$ integrals, we can transform \eqref{yyy} into a MB integral
\begin{align}\label{CI04first}
\CI_{0,4}&= \frac{2N \pi^4 h_0^4C_\phi^4 (2R)^{4\epsilon}\Gamma \left(\frac{\epsilon }{2}\right)^2}{\Gamma \left(1-\frac{\epsilon }{2}\right)^2 \Gamma (\epsilon )^2}\nonumber\\
&\times\int_{C_0}\frac{dz}{2\pi i}\frac{\Gamma \left(z\!-\!\frac{\epsilon }{2}\!+\!1\right) \Gamma
   \left(z+\frac{\epsilon }{2}\right)\Gamma (-z)^2   \Gamma \left(\frac{\epsilon }{2}-z\right)  \Gamma \left(\frac{3 \epsilon }{2}\!-\!1\!-\!z\right)}{ \Gamma (\epsilon
   -z)^2}    \,,
\end{align}

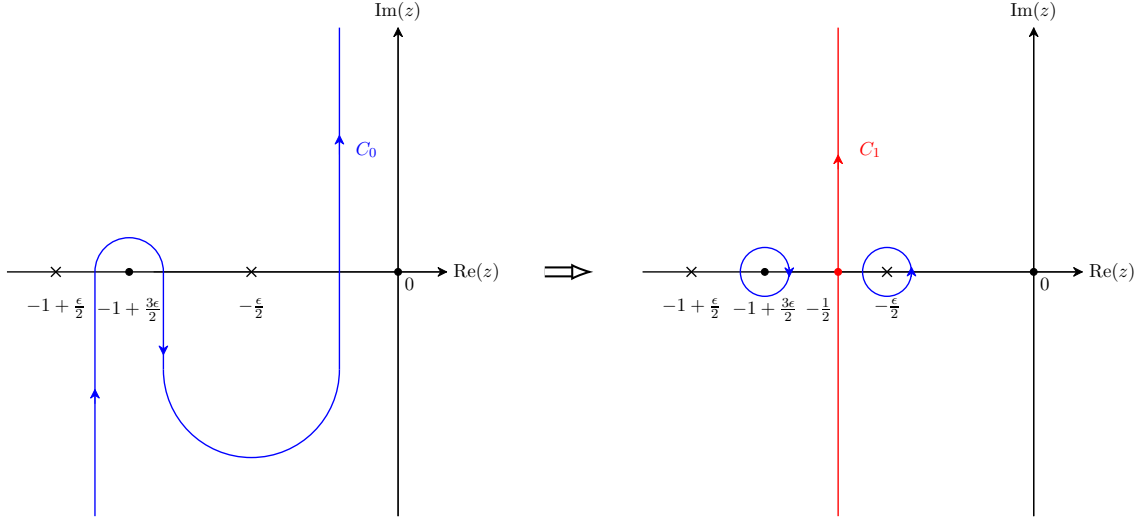
\begin{figure}
    \centering
  \scalebox{0.65}{
\begin{tikzpicture}
\pgfmathsetmacro\Ro{0.7}
\pgfmathsetmacro\Rt{1.8}
\pgfmathsetmacro\ep{3}
 \draw[  thick] (-8,0)--(1,0);
\draw [-{Stealth[round, length=6pt, width=6pt, bend]}](-5,0) -- (1,0) node[right] {$\text{Re}(z)$};
  \draw[  thick] (0,-5)--(0,5);
\draw [-{Stealth[round, length=6pt, width=6pt, bend]}](0,-5) -- (0,5) node[above] {$\text{Im}(z)$};

  \draw[thick,blue] (-6.2,-5) -- (-6.2,0); 
   \draw [-{Stealth[round, length=6pt, width=6pt, bend]},blue](-6.2,-2.7) -- (-6.2,-2.4);
  \draw[thick,blue] (-6.2,0) arc[start angle=180,end angle=0,radius=\Ro];
   \draw[thick,blue] (-4.8,0) -- (-4.8,-2); 
     \draw [-{Stealth[round, length=6pt, width=6pt, bend]},blue](-4.8,-1.5) -- (-4.8,-1.7);
   \draw[thick,blue] (-4.8,-2) arc[start angle=180,end angle=360,radius=\Rt];
   \draw[thick,blue] (-1.2,-2) -- (-1.2,5); 
     \draw [-{Stealth[round, length=6pt, width=6pt, bend]},blue](-1.2,2.4) -- (-1.2,2.8);
       \node[below right, blue] at (-1,2.8) {$C_0$};

   \foreach \x in {-7,-3} {
        \draw[thick] (\x-0.1,0.1) -- (\x+0.1,-0.1);
        \draw[thick] (\x-0.1,-0.1) -- (\x+0.1,0.1);
    }
    \node[below] at (-7,-0.4) {$-1+\frac{\epsilon}{2}$};
    \node[below] at (-3,-0.4) {$-\frac{\epsilon}{2}$};
      \filldraw[black] (-5.5,0) circle (2pt);
       \node[below] at (-5.5,-0.4) {$-1+\frac{3\epsilon}{2}$};
       \filldraw[black] (0,0) circle (2pt);
       \node[below right] at (0,0) {$0$};

        \draw[double distance=4pt, -{Latex[open]}, very thick] (3,0) -- (4,0);
        
\begin{scope}[xshift=13cm,yshift=0cm]
 \draw[  thick] (-8,0)--(1,0);
\draw [-{Stealth[round, length=6pt, width=6pt, bend]}](-5,0) -- (1,0) node[right] {$\text{Re}(z)$};
  \draw[  thick] (0,-5)--(0,5);
\draw [-{Stealth[round, length=6pt, width=6pt, bend]}](0,-5) -- (0,5) node[above] {$\text{Im}(z)$};

   \draw[thick,blue] (-5.5,0) circle(0.5);
    \draw [-{Stealth[round, length=6pt, width=6pt, bend]},blue](-5,0.1) -- (-5,-0.1);
 \draw[thick,blue] (-3,0) circle(0.5);
 \draw [-{Stealth[round, length=6pt, width=6pt, bend]},blue](-2.5,-0.1) -- (-2.5,0.1);

 \draw[thick,red] (-4,-5) -- (-4,5); 
 \draw [-{Stealth[round, length=6pt, width=6pt, bend]},red](-4,2.1) -- (-4,2.4);
  \node[below right, red] at (-3.7,2.8) {$C_1$};
   \filldraw[red] (-4,0) circle (2pt);
       \node[below left] at (-4,-0.4) {$-\frac{1}{2}$};

   \foreach \x in {-7,-3} {
        \draw[thick] (\x-0.1,0.1) -- (\x+0.1,-0.1);
        \draw[thick] (\x-0.1,-0.1) -- (\x+0.1,0.1);
    }
    \node[below] at (-7,-0.4) {$-1+\frac{\epsilon}{2}$};
    \node[below] at (-3,-0.4) {$-\frac{\epsilon}{2}$};
      \filldraw[black] (-5.5,0) circle (2pt);
       \node[below] at (-5.5,-0.4) {$-1+\frac{3\epsilon}{2}$};
       \filldraw[black] (0,0) circle (2pt);
       \node[below right] at (0,0) {$0$};
     
\end{scope}

\end{tikzpicture}
  }
    \caption{Contour deformation for the MB integral \eqref{CI04first}. The poles from the $\Gamma \left(z+\cdots\right)$ factors are labeled by crosses, while the poles from the $\Gamma\left(-z+\cdots\right)$ factors are labeled by black nodes. }
    \label{deformC0toC1}
\end{figure}
\noindent{}where the contour $C_0$ is shown in Figure \ref{deformC0toC1}. It separates the poles of all $\Gamma(z+...)$ factors from those of the $\Gamma(-z+...)$ factors. We deform the contour $C_0$ into a vertical line $C_1: \Re (z) = -\frac{1}{2}$. The integral along $C_1$ is well-defined in the $\epsilon\to0$ limit, while the divergent part of $\CI_{0,4}$ comes from the residues at the two poles picked up by the contour deformation (see Figure~\ref{deformC0toC1}),
\begin{align}
\CI_{0,4}&=2N \left[-5\frac{1+ \epsilon (2+2\gamma_E +4\log 2+2\log \pi )}{2^6 \pi ^4 \epsilon^2}+\CO(1)\right] h_0^4 R^{4\epsilon}\,~.
\end{align}

The second diagram $\CI_{1,3}$ involves one bulk integral and three defect integrals,
\begin{align}
\CI_{1,3} = \frac{2N(N+2)}{3}\lambda_0 h_0^3 C_\phi^5 R^{4\epsilon}\int \frac{d^d X_0 d^2 X_1 d^2 X_2 d^2 X_3}{X_{01}^{d-2}X_{02}^{d-2}X_{12}^{d-2}X_{03}^{2(d-2)}}\,.
\end{align}
We  integrate out  the bulk point using \eqref{Sdef0}, and the defect integral is of the form \eqref{CAdef},
\begin{align}
\CI_{1,3} 
& =  \frac{2N(N+2)}{3}\lambda_0 h_0^3 C_\phi^5 R^{4\epsilon}\nonumber\\
&\times\int \limits_{-i\infty}^{+i\infty} \frac{dz_1dz_2}{(2\pi i)^2}S_d\left(\frac{d-2}{2},\frac{d-2}{2},d-2; z_1, z_2\right) \CA_3\left(2d+z_1+z_2-5, -z_1, -z_2\right)\nonumber\\
&=\frac{2N(N+2)}{3}\left[-9\frac{ 1+\epsilon  (25/9+2 \gamma_E +4\log 2+2 \log \pi)}{2^{10} \pi ^5 \epsilon ^2}+\CO(1)\right]\lambda_0 h_0^3  R^{4\epsilon}\,,
\end{align}
where the explicit forms of $S_d$ and $\CA_3$ are given by \eqref{Sdef} and \eqref{CAfinal} respectively.

We proceed to compute the diagram $\CI_{2,2}^{(1)}$,
\begin{align}
\CI^{(1)}_{2,2} = \frac{N(N+2)}{6}\lambda_0^2 h_0^2 C_\phi^6 R^{4\epsilon} \int \frac{d^2 X_1 d^2 X_2 d^d X_3 d^d X_4}{X_{13}^{d-2}X_{14}^{d-2}X_{23}^{d-2}X_{24}^{d-2}X_{34}^{2(d-2)}}\,.
\end{align}
We integrate out the bulk points $(X_3, X_4)$ using \eqref{34MB}, and end up with an MB integral whose integrand is an integrated two-point function on $S^2$, which can be solved by \eqref{CAdef},
\begin{align}
\CI^{(1)}_{2,2} 
&= \frac{N(N+2)}{6}\lambda_0^2 h_0^2 C_\phi^6 R^{4\epsilon} \nonumber\\
&\times\int \limits_{-i\infty}^{+i\infty}\frac{dz_1dz_2}{(2\pi i)^2}S_d\left(\frac{d-2}{2},\frac{d-2}{2},d-2; z_1,z_2\right) C_{d;\frac{d-2}{2}-z_1, \frac{d-2}{2}-z_2}\CA_2\left(2(d-3)\right)\nonumber\\
&=\frac{N(N+2)}{6}\left[-\frac{1+\epsilon  (7/2+2 \gamma_E +4 \log2+2 \log \pi )}{2^{10} \pi ^6 \epsilon ^2}+\CO(1)\right]\lambda_0^2 h_0^2  R^{4\epsilon}\,.
\end{align}
Finally, the last two diagrams in Figure~\ref{4dia} can be evaluated easily using \eqref{Cdef} and \eqref{CAdef},
\begin{align}
\CI_{2,2}^{(2)} 
&=\frac{N(N+2)^2}{36}\lambda_0^2 h_0^2 C_\phi^6 R^{4\epsilon} C_{d;d-2, d-2}C_{d;\frac{3d}{2}-4, d-2}\CA_2\left(2(d-3)\right)
\nonumber\\
&=\frac{N(N+2)^2}{36}\left[-3\frac{ 1+\epsilon  (3+2 \gamma_E +4\log 2+2 \log \pi)}{2^{10} \pi ^6 \epsilon ^2}+\CO(1)\right]\lambda_0^2 h_0^2  R^{4\epsilon}\,,\nonumber\\
   \CI_{2,2}^{(3)} &=  \frac{N(N+2)}{9}\lambda_0^2 h_0^2 C_\phi^6 R^{4\epsilon} C_{d;\frac{d-2}{2}, 3\frac{d-2}{2}}C_{d;\frac{3d}{2}-4, \frac{d-2}{2}}\CA_2\left(2(d-3)\right)\\
   &= \frac{N(N+2)}{9}\left(\frac{1}{2^{13}\pi^6\epsilon}+\CO(1)\right) \lambda_0^2 h_0^2 R^{4\epsilon}\,. \nonumber
\end{align}
Altogether, the defect free energy to this order is 
\begin{align}
\CF= -\left(F_0+F_1+F_2+\CI_{0,4}+\, \CI_{1, 3}+\CI^{(1)}_{2,2}+ \CI^{(2)}_{2,2}+\CI^{(3)}_{2,2}\right)\,.
\end{align}

Keeping terms up to the quartic order in $\lambda$ and $h$ 
and plugging in the fixed points $\lambda_\star$ and $h_\star$, we obtain the $b$ anomaly for the ordinary surface defect in the IR,
\begin{align}
b = - 3 R\partial_R \CF = -\frac{27 N \epsilon ^3}{(N+8)^3}-\frac{27N(N+2)(52-N)\epsilon ^4}{4(N+8)^5}+\CO(\epsilon ^5)\,.
\label{beexp}
\end{align}
To leading order in the large $N$ limit, we have
\begin{align}
b \stackrel{N\to \infty }{=} -\frac{27\epsilon ^3}{N^2}+\frac{27\epsilon ^4}{4N^2}+\cdots \,,
\end{align}
which agrees with
(6.31) in \cite{Giombi:2023dqs} to the order of $\epsilon^4$.

For $d=3$, according to the factorization proposal, the $b$ anomaly of the ordinary surface is related to that of the ordinary boundary simply by
\ie 
b=2 b^{\rm bdry}.
\label{bbtwice}
\fe
In the Ising CFT, our result from $\ep$-expansion gives $b = -0.037$ with only the leading term, and $b = -0.054$ if we include the $\epsilon^4$ correction. In comparison, the fuzzy sphere result for the ordinary boundary of the Ising CFT is $b^{\rm bdry} = - 0.0159(5) $. Thus the leading order answer from $\ep$-expansion gives a better approximation (after taking into account the factor of two in \eqref{bbtwice}). This suggests that higher order corrections in the $\ep$-expansion are important and need to be resummed to give a more accurate prediction for $\ep=1$.\footnote{One can also attempt to extrapolate our results to $d=2$. For example, for $N=1$ and $\epsilon = 2$, our truncated $\ep$-expansion result \eqref{beexp} gives  $b \approx - 0.576$. In this case, the surface defect fills the whole spacetime and thus is equivalent to a mass deformation in the Ising CFT. The $b$ anomaly in this case is the difference between the IR and the UV conformal anomalies ($i.e.$ the usual $c$ anomaly).
It is well-known that the mass deformation gaps the Ising CFT, leading to $b = - \frac{1}{2}$ which agrees well with our naive extrapolation.}

\section*{Acknowledgements}
We thank Simone Giombi, Max Metlitski, Himanshu Khanchandani, Igor Klebanov, Fedor Popov and Yijian Zou for helpful discussions. 
ZS is supported in part by the US National Science Foundation
Grant No. PHY-2209997 and by the Simons Foundation Grant No. 917464. The work of YW is
supported in part by the NSF grant PHY-2210420 and by the Simons Junior Faculty Fellows program.

\appendix
\section{Assorted Integrals}\label{AI}
In this appendix, we list all integrals which are used in the main text. The main technique we employ is the Mellin-Barnes (MB) representation of the integrand for the multi-dimensional integrals that arise from the bulk-defect Feynman diagrams. One difficulty with these spacetime integrals has to do with the mixed integrations over the bulk and on the defect.
The MB method allows us to reduce such integrals to easier conventional massless spacetime integrals at the cost of introducing and integrating over a few MB variables. The latter MB integral can then be done either analytically or numerically with the help of the \texttt{Mathematica} package \texttt{MB.m} from \cite{Czakon:2005rk}.

\subsection{Integrals in flat space}
Let us start by discussing integrals we encounter in flat space with a flat surface defect.
The first integral is a well-known massless integral. It is a convolution of two conformal two-point functions in $n$ dimensions
\begin{align}\label{Cdef}
\int \frac{d^n X_3 }{X_{13}^{2a}X_{23}^{2b}} = \frac{C_{n; a, b}}{X_{12}^{2a+2b-n}}\,, \quad C_{n;a, b}\equiv  \pi^{\frac{n}{2}}\frac{\Gamma(n/2-a)\Gamma(n/2-b)\Gamma(a+b-n/2)}{\Gamma(a)\Gamma(b)\Gamma(n-a-b)}\,.
\end{align}
\normalsize
The second example is an integral of the product of three two-point functions 
\begin{align}\label{Sdef0}
    \CI_n(\gamma_1, \gamma_2, \gamma_3) = \int  \frac{d^n X_0}{X_{01}^{2\gamma_1}X_{02}^{2\gamma_2}x_{03}^{2\gamma_3}}\,,
\end{align}
which does not appear to admit analytical evaluation for generic $\gamma_i$. Here we aim to provide a MB representation of this integral. The derivation is standard and can be applied to many other examples. 

We start with applying Feynman parametrization to the three terms in the denominator and then perform the $X_0$ integral
\begin{align}\label{CIgamma}
    \CI_n(\gamma_1, \gamma_2, \gamma_3) &= \frac{\Gamma(\gamma_1+\gamma_2+\gamma_3)}{\Gamma(\gamma_1)\Gamma(\gamma_2)\Gamma(\gamma_3)}\int_0^1 d^3 u \, \delta(1-\sum_i u_i)\int d^n X_0\frac{u_1^{\gamma_1-1}u_2^{\gamma_2-1}u_3^{\gamma_3-1}}{(u_1 X_{01}^2+u_2 X_{02}^2+u_3 X_{03}^2)^{\sum_i\gamma_i}}\nonumber\\
    &=\frac{\pi^{\frac{n}{2}}\Gamma(\sum_i\gamma_i-\frac{n}{2})}{\Gamma(\gamma_1)\Gamma(\gamma_2)\Gamma(\gamma_3)}\int_0^1 d^3 u\frac{\delta(1-\sum_i u_i)u_1^{\gamma_1-1}u_2^{\gamma_2-1}u_3^{\gamma_3-1}}{(u_1 u_2 X_{12}^2+u_1 u_3 X_{13}^2+u_2 u_3 X_{23}^2)^{\sum_i\gamma_i-\frac{n}{2}}}\,.
\end{align}
To evaluate the remaining integral over the Feynman parameters, we implement the MB representation,
\begin{align}\label{basicMB}
    \frac{\Gamma(\alpha)}{(A_1+A_2+\cdots A_{K+1})^\alpha} = \int \limits_{-i\infty}^{+i\infty}\prod_{j=1}^K \left(\frac{dz_j}{2\pi i}\Gamma(-z_j)A_j^{z_j}\right)\frac{\Gamma(z_1+\cdots+ z_K+\alpha)}{A_{K+1}^{z_1+\cdots+z_K+\alpha}}\,,
\end{align}
and use the following well-known integral
\begin{align}\label{us}
    \int_0^1 d^L u \, \delta(1-u_1-\cdots-u_L) u_1^{a_1-1}\cdots u_L^{a_L-1} = \frac{\Gamma(a_1)\cdots \Gamma(a_L)}{\Gamma(a_1+\cdots +a_L)}\,.
\end{align}
Plugging \eqref{basicMB} and \eqref{us} into \eqref{CIgamma} yields 
\begin{align}\label{SSd}
  \CI_n(\gamma_1, \gamma_2, \gamma_3) = \int\limits_{-i\infty}^{+i\infty}\frac{dz_1dz_2}{(2\pi i)^2} S_n(\gamma_1,\gamma_2,\gamma_3; z_1,z_2) \frac{X_{13}^{2z_1}X_{23}^{2z_2}}{X_{12}^{2(z_1+z_2+\gamma_1+\gamma_2+\gamma_3)-n}}\,,
\end{align}
where 
\begin{align}\label{Sdef}
 S_n(\gamma_1,\gamma_2,\gamma_3; z_1,z_2)&=\pi^{\frac{n}{2}}\frac{\Gamma(-z_1)\Gamma(-z_2)\Gamma(\frac{n}{2}\!-\!\gamma_1\!-\!\gamma_3\!-\!z_1)\Gamma(\frac{n}{2}\!-\!\gamma_2\!-\!\gamma_3\!-\!z_2)}{\Gamma(\gamma_1)\Gamma(\gamma_2)\Gamma(\gamma_3)}\nonumber\\
 &\times \frac{\Gamma(\gamma_3\!+\!z_1\!+\!z_2)\Gamma(\gamma_1+\gamma_2+\gamma_3\!+\!z_1\!+\!z_2-\frac{n}{2})}{\Gamma(n- \gamma_1-\gamma_2-\gamma_3)}\,.
\end{align}
Equation \eqref{SSd} serves as a useful MB representation of the integral $\CI_n(\gamma_1, \gamma_2, \gamma_3)$ over one bulk point. 

For the following integral over two bulk points, 
we obtain after combining \eqref{Cdef} and \eqref{SSd}, 
\begin{align}\label{34MB}
\int  \frac{d^n X_3 d^n X_4}{\prod_{1\le i<j\le 4}X_{ij}^{2\gamma_{ij}}} = \left(X^2_{12}\right)^{n-\sum_{i<j}\gamma_{ij}} \int\limits_{-i\infty}^{+i\infty}\frac{dz_1 dz_2}{(2\pi i)^2}S_n(\gamma_{14},\gamma_{24},\gamma_{34}; z_1,z_2) C_{n;\gamma_{13}-z_1, \gamma_{23}-z_2}\,.
\end{align} 

The last example we discuss here is an integral in flat space that does not have the 
full translation symmetry (the integrand is a product of two bulk-defect two-point functions)
\begin{align}
\CB (a_1, a_2, b) = x_{12}^{2(a_1+a_2+b)-d}\int \frac{d^2 x_3 d^{d-2}\eta}{(x_{13}^2+\eta^2)^{a_1}(x_{23}^2+\eta^2)^{a_2}\eta^{2b}}\,.
\end{align}
We apply Schwinger parametrization to the first two terms in the denominator. This allows us to integrate out $x_3$ and $\eta$ easily
\begin{align}
\CB (a_1, a_2, b) = x_{12}^{2(a_1+a_2+b)-d}\frac{\pi\text{Vol}(S^{d-3})\Gamma(\frac{d}{2}-b-1)}{2\Gamma(a_1)\Gamma(a_2)}\int_0^\infty\frac{ds_1ds_2}{s_1s_2}\frac{s^{a_1}_1 s_2^{a_2}}{(s_1+s_2)^{\frac{d}{2}-b}} e^{-\frac{s_1 s_2}{s_1+s_2}x_{12}^2}\,.
\end{align}
Then we make the rescaling $s_2\to s_1 s_2$ and integrate out $s_1$ (the $x_{12}$ dependence drops out for obvious reasons)
\begin{align}\label{CB}
\CB (a_1, a_2, b) &= \frac{\pi\text{Vol}(S^{d-3})\Gamma(\frac{d}{2}-b-1)}{2\Gamma(a_1)\Gamma(a_2)}\int_0^\infty\frac{ds_1ds_2}{s_1s_2}\frac{s^{a_1+a_2+b-\frac{d}{2}}_1 s_2^{a_2}}{(1+s_2)^{\frac{d}{2}-b}} e^{-\frac{s_1 s_2}{1+s_2}}\\
&= \frac{\pi\text{Vol}(S^{d-3})\Gamma(\frac{d}{2}-b-1)\Gamma(a_1+a_2+b-\frac{d}{2})}{2\Gamma(a_1)\Gamma(a_2)}\int_0^\infty\frac{ds_2}{s_2^{1+a_1+b-\frac{d}{2}}(1+s_2)^{d-a_1-a_2-2b}}\nonumber\\
&= \frac{\pi\text{Vol}(S^{d-3})\Gamma(\frac{d}{2}-b-1)\Gamma(\frac{d}{2}-b-a_1)\Gamma(\frac{d}{2}-b-a_2)\Gamma(a_1+a_2+b-\frac{d}{2})}{2\Gamma(a_1)\Gamma(a_2)\Gamma(d-a_1-a_2-2b)}\,,\nonumber
\end{align}
where $\text{Vol}(S^{n}) = \frac{2 \pi ^{\frac{n+1}{2}}}{\Gamma \left(\frac{n+1}{2}\right)}$ is the volume of a unit $n$-sphere.

\subsection{Integrals on sphere}
In the remaining part of this appendix, we consider integrals of two- and three-point functions over $S^2$ which is relevant for a spherical surface defect,
\begin{align}\label{CAdef}
\CA_2(\gamma) \equiv  \int_{S^2}\frac{d^2 X_1 d^2 X_2}{X_{12}^{2\gamma}}\,, \quad \CA_3(\gamma_{12}, \gamma_{13},\gamma_{23}) \equiv \int_{S^2} \frac{d^2 X_1 d^2 X_2 d^2 X_3}{X_{12}^{2\gamma_{12}}X_{13}^{2\gamma_{13}}X_{23}^{2\gamma_{23}}}\,,
\end{align}
where the notations $d^2 X$ and $X^2_{ij}$ are explained in Section \ref{nac}.
 $\CA_2$ can be found in \cite{Cardy:1988cwa},
\begin{align}\label{CardyA2}
\CA_2(\gamma) = \frac{4^{2-\gamma }\pi ^2 }{1-\gamma }\,.
\end{align}
For $\CA_3$, we write it in stereographic coordinates which are reviewed in Section \ref{nac}, and then use the rotation symmetry to fix $y_3 = 0$:
\begin{align}
\CA_3(\gamma_{12}, \gamma_{13},\gamma_{23}) =4\pi  \int\frac{d^2y_1\Omega_1^2 d^2y_2 \Omega_2^2}{(\Omega_1\Omega_2 y_{12}^2)^{\gamma_{12}}(2\Omega_1 y_1^2)^{\gamma_{13}} (2\Omega_2 y_2^2)^{\gamma_{23}}}\,.
\end{align}
The integrand can be further simplified by performing inversions of $y_1$ and $y_2$
\begin{align}
\CA_3(\gamma_{12}, \gamma_{13},\gamma_{23}) &= 4^{3-\gamma_{12}-\gamma_{13}-\gamma_{23}}\pi\int \frac{d^2 y_1 d^2 y_2}{(1+y_1^2)^{2-\gamma_{12}-\gamma_{13}}(1+y_2^2)^{2-\gamma_{12}-\gamma_{23}}y_{12}^{2\gamma_{12}}}\,.
\end{align}
The remaining integral is a special case of \cite{Fei:2015oha}
\begin{align}\label{G0int}
    \int \frac{d^n y_1 d^n y_2}{(1+y_1^2)^{a_1}(1+y_2^2)^{a_2} y_{12}^{2b}} = \Gamma_0(n; a_1, a_2, b)\,,
\end{align}
where
\begin{align}\label{G0}
\Gamma_0(n; a_1, a_2, b)=\frac{\pi^n\Gamma(\frac{n}{2}-b)\Gamma(a_1+b-\frac{n}{2})\Gamma(a_2+b-\frac{n}{2})\Gamma(a_1+a_2+b-n)}{\Gamma(\frac{n}{2})\Gamma(a_1)\Gamma(a_2)\Gamma(a_1+a_2+2b-n)}\,.
\end{align}
Therefore, $\CA_3$ can be compactly expressed as 
\begin{align}\label{CAfinal}
\CA_3(\gamma_{12}, \gamma_{13},\gamma_{23}) =4^{3-\gamma_{12}-\gamma_{13}-\gamma_{23}}\pi\Gamma_0\left(2; 2-\gamma_{12}-\gamma_{13},2-\gamma_{12}-\gamma_{23}, \gamma_{12}\right)\,.
\end{align}
When $\gamma_{13}=\gamma_{23}=0$, it reduces to $4\pi\CA_2(\gamma_{12})$.

Another useful integral from Appendix B in \cite{Fei:2015oha} is the following,
\begin{gather}
\label{G1int}
    \int \frac{d^ny}{(1+y^2)^{a}(x-y)^{2b}}=\int\limits_{-i\infty}^{+i\infty}\frac{dz_1}{2\pi i}\frac{\Gamma_{1}\left(n;a,b,z_1\right)}{(1+x^2)^{-z_1}}\,,
\end{gather}
where
\begin{gather}
\label{G1}
    \Gamma_1(n;a,b,z_1)=\frac{\pi^{n/2} \Gamma(-z_1) \Gamma\left(a+b-\frac{n}{2}+z_1\right)\Gamma\left(b+z_1\right)\Gamma\left(n-a-2b-z_1\right)}{\Gamma (a) \Gamma(b)\Gamma (n-a-b)}\,.
\end{gather}

\section{Normalization of  the Bulk Local Operators}\label{Z2}

In this appendix, we compute the renormalized  two-point functions $\langle \phi^{I}(X_1)\phi^{I}(X_2) \rangle$ (no sum over $I$)  and $\langle[\phi^2]_R(X_1)[\phi^2]_R(X_2)\rangle$ where $\phi^2\equiv \phi^{J}\phi^{J}$, at the WF fixed point \eqref{fixedlambda}
of the $O(N)$ model \eqref{SON}. We determine the overall normalization of these two-point functions. Our results hold up to the $\epsilon^2$ order at the fixed point.

We start with the bare $\phi^I_0$ operator. To order $\lambda_0^2$, the corresponding bare two-point function takes the following form
\begin{align}
    \langle \phi_0^{I}(X_1)\phi_0^{I}(X_2) \rangle=G_{d}(X_1,X_2)+G^{(2)}_{d}(X_1,X_2)\,,
\end{align}
where 
\begin{gather}
    G^{(2)}_{d}(X_1,X_2)=\frac{N+2}{18}  C^{4}_{\phi}C_{d;\frac{d-2}{2},\frac{3(d-2)}{2}}C_{d;\frac{d-2}{2},\frac{3d-8}{2}}\lambda_0^2 G_d(X_1, X_2)X_{12}^{2\epsilon}\,.
\end{gather}
By imposing the finiteness of $Z^{-1}_\phi \langle \phi_0^{I}(X_1)\phi_0^{I}(X_2) \rangle$ we recover  the wavefunction renormalization $Z_{\phi}$,
\begin{gather}\label{Zphi}
    Z_{\phi}=1-\frac{N+2}{3}\frac{\lambda^{2}}{(4\pi)^{4}}\frac{1}{12 \epsilon}\,,
\end{gather}
which gives the anomalous dimension of $\phi^{I}$ at the WF fixed point $\lambda_\star$:
\begin{gather}
    \gamma_{\phi}=\frac{1}{2}\beta_{\lambda}\partial_\lambda\log Z_{\phi}\big|_{\lambda_\star}=\frac{N+2}{36}\frac{\lambda^2}{(4\pi)^4}\bigg|_{\lambda\star}=\frac{ (N+2)}{4 (N+8)^2}\epsilon^2\,.
\end{gather}
We also provide the full two-point function of $\phi^{I}$ at the fixed point $
\lambda_\star$
\begin{gather}\label{phiphinodefect}
    \langle \phi^I(X_1)\phi^I(X_2)\rangle=\dfrac{\CN^2_{\phi}}{X_{12}^{2\left(\frac{d-2}{2}+\gamma_{\phi}\right)}}\,, 
\end{gather}
where the normalization is 
\begin{gather}
    \CN^2_{\phi}=\frac{1}{4 \pi ^2}+\frac{\epsilon \log (\pi e^{\gamma_E} )}{8 \pi ^2} 
    -\frac{\epsilon^2}{192 \pi ^2 (N+8)^2} \bigg(N \left(39-\pi ^2 (N+16)\right)\nonumber \\-6 (N+8)^2 \log ^2(\pi e^{\gamma_E})+12 (N+2) \log ( \pi e^{\gamma_E} )-64 \pi ^2+78\bigg)\,.
    \label{CNphibulk}
\end{gather}

\begin{figure}[t]
\centering
\begin{tikzpicture} 
\draw[color=black] (0,0) circle [radius=0.6]; 
\draw (-0.6,0)node[vertex]{} ;
\draw (0.6,0)node[vertex]{} ;
\node at (0,-1) {$\CG_0$};
\end{tikzpicture} 
\qquad
\begin{tikzpicture} 
\draw[color=black] (0,0) circle [radius=0.6]; 
\draw[color=black] (1.2,0) circle [radius=0.6]; 
\draw (-0.6,0)node[vertex]{} ;
\draw (0.6,0)node[vertex]{} ;
\draw (1.8,0)node[vertex]{} ;
\node at (0.6,-1) {$\CG_1$};
\end{tikzpicture} 
\qquad
\begin{tikzpicture} 
\draw[color=black] (0,0) circle [radius=0.6]; 
\draw[color=black] (1.2,0) circle [radius=0.6]; 
\draw[color=black] (2.4,0) circle [radius=0.6]; 
\draw  (-0.6,0)node[vertex]{} ;
\draw (0.6,0)node[vertex]{} ;
\draw (1.8,0)node[vertex]{} ;
\draw (3,0)node[vertex]{} ;
\node at (1.2,-1) {$\CG_{21}$};
\end{tikzpicture} 
\qquad
\begin{tikzpicture} 
\draw[color=black] (0,0) circle [radius=0.6]; 
\draw(-0.6,0)node[vertex]{} ;
\draw (0.6,0)node[vertex]{} ;
\draw (-0.4, 0.45)node[vertex]{} to  (0.4,0.45)node[vertex]{} ;
\draw (-0.4, 0.45) to[out=-45,in=-135]    (0.4,0.45) ;
\node at (0,-1) {$\CG_{22}$};
\end{tikzpicture} 
\qquad
\begin{tikzpicture} 
\draw[color=black] (0,0) circle [radius=0.6]; 
\draw(-0.6,0)node[vertex]{} ;
\draw (0.6,0)node[vertex]{} ;
\draw (0,0.6)node[vertex]{} to [out=-135,in=135]  (0,-0.6)node[vertex]{} ;
\draw (0,0.6)node[vertex]{} to [out=-45,in=45]  (0,-0.6)node[vertex]{} ;
\node at (0,-1) {$\CG_{23}$};
\end{tikzpicture} 
\caption{The two-point function of $\phi^2$ in $O(N)$ model up to three loops.}
\label{phiphi}
\end{figure}
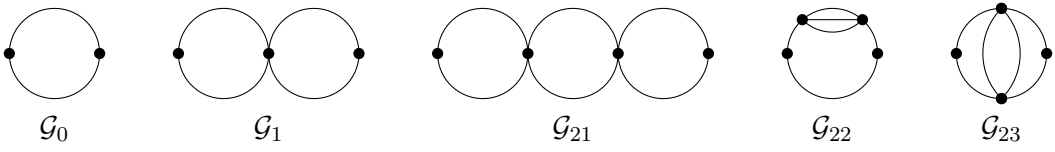

Next, we consider the bare $\phi_0^2=\phi^I_0\phi^I_0$ operator. To the order of $\lambda_0^2$, the relevant diagrams are shown in Figure \ref{phiphi}, i.e.
\begin{align}
    \left\langle\phi_0^2(X_1)\phi_0^2(X_2)\right\rangle = \CG_0 +\CG_1+\CG_{21}+\CG_{22}+\CG_{23}\,.
\end{align}
The first diagram $\CG_0$ is simply the product of two free propagators, multiplied by a factor of $2N$.
The following three diagrams admit simple analytic expressions:
\begin{align}
&\CG_1(X_1, X_2) =  -\frac{N+2}{6}\lambda_0 C_\phi^2 C_{d; d-2, d-2}\CG_0(X_1, X_2) X_{12}^\epsilon\,,\nonumber\\
&\CG_{21}(X_1, X_2) =\left(\frac{N+2}{6}\lambda_0\right)^2 C_\phi^4 C_{d;d-2, d-2} C_{d;d-2, \frac{3d-8}{2}} \CG_0(X_1, X_2)X_{12}^{2\epsilon}\,,\\
&\CG_{22}(X_1, X_2) = \frac{N+2}{9}\lambda_0^2 C_\phi^4C_{d;\frac{d-2}{2},3\frac{d-2}{2}}C_{d;\frac{d-2}{2}, \frac{3d-8}{2}} \CG_0(X_1, X_2)X_{12}^{2\epsilon}\,.\nonumber
\end{align}
which can be derived by using \eqref{Cdef} repeatedly.
We don't aim to analytically computing the last diagram. Instead we can straightforwardly rewrite it as a double MB integral using \eqref{34MB}, and then extract the $\epsilon$-expansion. In order to obtain the precise normalization of $\langle[\phi^2]_R(X_1)[\phi^2]_R(X_2)\rangle$ , we have to keep the order $\epsilon^0$ terms in the expansion:
\begin{align}
\CG_{23} =\frac{N+2}{24}\lambda_0^2 C_\phi^2\CG_0(X_1, X_2)X_{12}^{2\epsilon}\left[\frac{1}{\epsilon ^2}+\frac{3}{2\epsilon}+\frac{3}{4}+\CO(\epsilon)\right]\,.
\end{align}
Imposing the finiteness of $Z^{-2}_{\phi^2}\langle \phi_0^2(X_1)\phi_0^2(X_2)\rangle$ and using \eqref{lambdabare}, we recover the wavefunction renormalization $Z_{\phi^2}$ to order $\lambda^2$:
\begin{align}
Z_{\phi^2}=1-\frac{(N+2)\lambda }{3(4\pi) ^2 \epsilon }+\frac{5(N+2) \lambda ^2}{36 (4\pi) ^4 \epsilon }-\frac{(N+2)\lambda ^2}{3(4\pi) ^4 \epsilon ^2}+\CO(\lambda^3)\,,
\end{align} 
which yields the anomalous dimension of $\phi^2$ at the WF fixed point $\lambda_\star$:
\begin{align}\label{gammaphi2}
\gamma_{\phi^2} = \left.\beta_\lambda\partial_\lambda\log Z_{\phi^2}\right|_{\lambda_\star} = \frac{N+2 }{N+8} \epsilon+\frac{(N+2) (13 N+44)}{2 (N+8)^3} \epsilon ^2+\CO(\epsilon^3)\,.
\end{align}
We also report the full two-point function of $[\phi^2]_R$ at the WF fixed point
\begin{align}
\left\langle [\phi^2]_R(X_1)[\phi^2]_R(X_2)\right\rangle  = \frac{\CN_{\phi^2}^2}{X_{12}^{2(d-2+\gamma_{\phi^2})}}\,,
\end{align} 
where
\begin{align}\label{phi2normalization}
\CN_{\phi^2}^2&= 2N C_\phi^2 \left[1-\frac{(N+2)(1+\log(\pi e^{\gamma_E}))}{N+8}\epsilon\right]+N C_\phi^2\left[\frac{(N+2)\log(\pi e^{\gamma_E})}{(N+8)}\epsilon\right]^2\\
& \,\,+N (N+2)C_\phi^2\left[\frac{(2N^2+7N-12)\log(\pi e^{\gamma_E})}{(N+8)^3}-\frac{ \left(\pi ^2 (N+8)^2+147 N+456\right) }{12 (N+8)^3}\right]\epsilon ^2+\CO(\epsilon^3)\,.\nonumber
\end{align}
The order $\epsilon$ term agrees with \cite{Cuomo:2021kfm}.
\section{The scaling dimension of $[\hat\phi^2]_R$ to the $\epsilon^3$ order}\label{ordereps3}
In this appendix, we compute the fixed point $h_{\star}$ and the scaling dimension of the defect operator $[\hat\phi^2]_R$ to the  $\epsilon^3$ order. 

\begin{figure}[!htb]
\centering
\begin{tikzpicture} 
\draw [line width = 0.5mm] (-1.2, -1) to (1.2,-1);
\draw   (0,1)node[vertex]{} to  (-1,-1)node[vertex]{}  to [out=45,in=135]  (-0.34,-1)node[vertex]{}  to  [out=45,in=135] (0.32,-1) node[vertex]{}  to  [out=45,in=135] (1,-1) node[vertex]{} to (0,1);
\node at (0.1,-1.5) {$\CJ_{0,4}$};
\end{tikzpicture}
\quad 
\begin{tikzpicture} 
\draw [line width = 0.5mm] (-1.2, -1) to (1.2,-1);
\draw   (0,1) node[vertex]{} to[out=-45,in=45]  (0,0)node[vertex]{} to[out=135,in=-135]  (0,1);
\draw   (0,0) to (-1,-1)node[vertex]{} to [out=45,in=135] (0,-1)node[vertex]{}  to [out=45,in=135] (1,-1) node[vertex]{} to (0,0)  ; 
\node at (0.1,-1.5) {$\CJ^{(1)}_{1,3}$};
\end{tikzpicture}
\qquad 
\begin{tikzpicture} 
\draw [line width = 0.5mm] (-1.2, -1) to (1.2,-1);
\draw   (0,1) node[vertex]{} to (-0.5,0)node[vertex]{} to  (-1,-1)node[vertex]{};
\draw (-0.5,0)node[vertex]{} to  (0,-1)node[vertex]{};
\draw (-1,-1)node[vertex]{} to[out=45,in=135]  (0,-1)node[vertex]{};
\draw  (-0.5,0)   to (1,-1);
\draw   (0,1) to (1,-1) node[vertex]{} ; 
\node at (0.1,-1.5) {$\CJ^{(2)}_{1,3}$};
\end{tikzpicture}
\quad
\begin{tikzpicture} 
\draw [line width = 0.5mm] (-1.2, -1) to (1.2,-1);
\draw   (0,1) node[vertex]{} to (-0.5,0)node[vertex]{} to[out=-140,in=80]  (-1,-1)node[vertex]{};
\draw  (-0.5,0) to[out=-100,in=40]  (-1,-1)node[vertex]{};
\draw  (-0.5,0)   to (0,-1) node[vertex]{};
\draw   (0,1) to (1,-1) node[vertex]{}to [out=135,in=45] (0,-1); 
\node at (0.1,-1.5) {$\CJ^{(3)}_{1,3}$};
\end{tikzpicture}
\quad
\begin{tikzpicture} 
\draw [line width = 0.5mm] (-1.2, -1) to (1.2,-1);
\draw   (0,1) node[vertex]{} to  (-1,-1)node[vertex]{} to  (0,0)node[vertex]{} to  (1,-1)node[vertex]{} to (0,1);
\draw (0,0) node[vertex]{} to[out=-45,in=45]  (0,-1)node[vertex]{} to[out=135,in=-135]   (0,0);
\node at (0.1,-1.5) {$\CJ_{1,3}^{(4)}$};
\end{tikzpicture}
\\
\begin{tikzpicture} 
\draw [line width = 0.5mm] (-1.2, -1) to (1.2,-1);
\draw   (0,1) node[vertex]{} to[out=-45,in=45]  (0,0.5) node[vertex]{} to[out=-45,in=45] (0,0)node[vertex]{} to[out=135,in=-135]  (0,0.5) node[vertex]{} to[out=135,in=-135] (0,1);
\draw   (0,0) to (-1,-1)node[vertex]{} to [out=45,in=135] (1,-1)node[vertex]{} to (0,0)  ; 
\node at (0.1,-1.5) {$\CJ^{(1)}_{2,2}$};
\end{tikzpicture}
\quad
\begin{tikzpicture} 
\draw [line width = 0.5mm] (-1.2, -1) to (1.2,-1);
\draw   (0,1) node[vertex]{} to[out=-45,in=45]  (0,0.5)node[vertex]{} to[out=135,in=-135]  (0,1);
\draw   (0,0.5) node[vertex]{} to (-0.5,0)node[vertex]{} to[out=-140,in=80]  (-1,-1)node[vertex]{};
\draw  (-0.5,0) to[out=-100,in=40]  (-1,-1)node[vertex]{};
\draw  (-0.5,0)   to (1,-1);
\draw   (0,0.5) to (1,-1) node[vertex]{}; 
\node at (0.1,-1.5) {$\CJ^{(2)}_{2,2}$};
\end{tikzpicture}
\quad
\begin{tikzpicture} 
\draw [line width = 0.5mm] (-1.2, -1) to (1.2,-1);
\draw   (0,1) node[vertex]{} to (-0.5,0)node[vertex]{} to[out=-140,in=80]  (-0.75,-0.5)node[vertex]{} to[out=-140,in=80](-1,-1)node[vertex]{};
\draw  (-0.5,0) to[out=-100,in=40]  (-0.75,-0.5)node[vertex]{} to[out=-100,in=40]  (-1,-1);
\draw  (-0.5,0)   to (1,-1);
\draw   (0,1) to (1,-1) node[vertex]{}; 
\node at (0.1,-1.5) {$\CJ^{(3)}_{2,2}$};
\end{tikzpicture}
\quad
\begin{tikzpicture} 
\draw [line width = 0.5mm] (-1.2, -1) to (1.2,-1);
\draw   (0,1) node[vertex]{} to (-0.5,0)node[vertex]{} to[out=-140,in=80]  (-0.75,-0.5)node[vertex]{} to(-1,-1)node[vertex]{};
\draw  (-0.5,0) to[out=-100,in=40]  (-0.75,-0.5)node[vertex]{} to  (-1,-1);
\draw  (-1,-1) to[out=135,in=-200]  (-0.5,0);
\draw  (-0.75,-0.5)   to (1,-1);
\draw   (0,1) to (1,-1) node[vertex]{}; 
\node at (0.1,-1.5) {$\CJ^{(4)}_{2,2}$};
\end{tikzpicture}
\\
\begin{tikzpicture} 
\draw[line width = 0.5mm] (-1.2, -1) to (1.2,-1);
\draw   (0,1)node[vertex]{} to  (-1,-1)node[vertex]{}  to [out=45,in=180]  (-0.25,-0.6);
\draw  (0.25,-0.6)  to  [out=0,in=135] (1,-1)node[vertex]{} to (0,1);
\draw  (-0.25,-0.6) to (0.25,-0.6);
\node at (0.1,-1.5) {$\mathcal{J}^{(5)}_{2,2}$};

\draw (-0.25,-0.6) node[]{} to [out=45,in=135] (0.25,-0.6) node[]{} [out=-135,in=-45] to (-0.25,-0.6);
\draw (-0.25,-0.6) node[vertex]{} ;
\draw (0.25,-0.6) node[vertex]{} ;
\end{tikzpicture}
\quad
\begin{tikzpicture} 
\draw [line width = 0.5mm] (-1.2, -1) to (1.2,-1);
\draw   (0,1)node[vertex]{} to  (-1,-1)node[vertex]{}  to [out=45,in=135]  (1,-1)node[vertex]{}  to  [out=45,in=135] (1,-1)node[vertex]{} to (0.6,-0.2);
\draw  (0,1) to (0.4,0.2);

\draw [rotate=-63.43, shift={(0.2,1.05)}] 
    (-0.25,-0.6) node[]{} 
    to [out=45,in=135] (0.25,-0.6) node[]{} 
    [out=-135,in=-45] 
    to (-0.25,-0.6); 

\draw (0.4,0.2) node[vertex]{} ;
\draw (0.6,-0.2) node[vertex]{} ;

\draw (0.6,-0.2) to (0.4,0.2);

\node at (0.1,-1.5) {$\mathcal{J}^{(6)}_{2,2}$};
\end{tikzpicture}
\quad
\begin{tikzpicture} 
\draw [line width = 0.5mm] (-1.1, -1) to (1.1,-1);

\draw   (0,1) node[vertex]{} to  (-0.8,0)node[vertex]{} to (0.8,0) node[vertex]{} to  (0,1) ;
\draw (-0.8,0) node[vertex]{} to[out=-55,in=55]  (-0.8,-1)node[vertex]{} to[out=125,in=-125]   (-0.8,0);
\draw (0.8,0) node[vertex]{} to[out=-55,in=55]  (0.8,-1)node[vertex]{} to[out=125,in=-125]   (0.8,0);
\node at (0.1,-1.5) {$\CJ_{2,2}^{(7)}$};
\end{tikzpicture}
\quad
\begin{tikzpicture} 
\draw [line width = 0.5mm] (-1.1, -1) to (1.1,-1);

\draw   (0,1) node[vertex]{} to  (-0.8,0)node[vertex]{} to (0.8,0) node[vertex]{} to  (0,1) ;
\draw (-0.8,0) node[vertex]{} to (-0.8,-1);
\draw (0.8,0) node[vertex]{} to (0.8,-1)node[vertex]{};
\draw (-0.8,-1) node[vertex]{} to[out=45,in=135]  (0.8,-1);
\draw (-0.8,0) node[vertex]{} to[out=-45,in=-135]  (0.8,0);
\node at (0.1,-1.5) {$\CJ_{2,2}^{(8)}$};
\end{tikzpicture}
\quad
\begin{tikzpicture} 
\draw [line width = 0.5mm] (-1.1, -1) to (1.1,-1);

\draw   (0,1) node[vertex]{} to  (-0.8,0)node[vertex]{} to (0.8,0) node[vertex]{} to  (0,1) ;
\draw (-0.8,0) node[vertex]{} to (-0.8,-1)node[vertex]{};
\draw (0.8,0) node[vertex]{} to (0.8,-1)node[vertex]{};
\draw (-0.8,-1) to (0.8,0);
\draw (-0.8,-1) to (0.8,0);
\draw (0.8,-1) to (0.1,-0.5625);
\draw (-0.1,-0.4375) to (-0.8,0);
\node at (0.1,-1.5) {$\CJ_{2,2}^{(9)}$};
\end{tikzpicture}
\caption{Diagrams contributing to $\langle \phi^2(\eta)\rangle$ at orders $h_0^4$, $h_0^3\lambda_0$, and $h_0^2\lambda_0^2$. }
\label{4loopdiag}
\end{figure}
Altogether, the  
one-point function of the renormalized bulk   operator $[\phi^2]_R = Z_{\phi^2}^{-1} \phi_0^2$, with contributions up to the fourth order in coupling constants reads
\begin{gather}
\left\langle [\phi^2]_{R}\right\rangle_{4-\text{order}} = \frac{1}{Z_{\phi^2}}\bigg(A_0+A_1+B+\CJ_{0,3}+\CJ^{(1)}_{1,2}+\CJ^{(2)}_{1,2}+ \CJ_{2,1}^{(1)}+\CJ_{2,1}^{(2)}+\CJ_{2,1}^{(3)}\nonumber\\
+16 N h^4_{0}\CJ^{}_{0,4}+\lambda_0 h^3_0\sum \limits_{\alpha=1}^{4}a^{(\alpha)}_{1,3}\CJ^{(\alpha)}_{1,3}+\lambda^2_0 h^2_0\sum \limits_{\alpha=1}^{9}a^{(\alpha)}_{2,2}\CJ^{(\alpha)}_{2,2}+ h_{0}\lambda^3_0 \CJ_{3,1}\bigg)\,,
\label{1ptordeps3}
\end{gather}
The diagrams that produce $A_0$ to $\CJ_{2,1}^{(3)}$ in the above are already provided in Section~\ref{MassOperator},\footnote{Among these diagrams, the full analytic expressions of $\CJ_{1,2}^{(1)}, \CJ_{1,2}^{(2)}$ and $\CJ_{2,1}^{(1)}$ are not known. We have obtained the divergent part of  their $\epsilon$-expansions in Section~\ref{MassOperator}. In this appendix, we also need their finite parts. Such data can be extracted easily with the help of the package \texttt{MB.m} \cite{Czakon:2005rk} given the MB representation of these three diagrams.} while the diagrams at orders $h_0^4$, $h_0^3\lambda_0$, and $h_0^2\lambda_0^2$ are shown in Figure \ref{4loopdiag}. The symmetry factors $a^{(\beta)}_{i,j}$ for each of the diagrams $\CJ^{(\beta)}_{i,j}$ are listed below,
\begin{align}
&a^{(1)}_{1,3}=a^{(4)}_{1,3}=\frac{4N(N+2)}{3}\,,\quad a^{(2)}_{1,3}=a^{(3)}_{1,3}=\frac{8N(N+2)}{3}\,,\nonumber\\
    &a_{2,2}^{(1)}= \frac{N(N+2)^2}{9}\,,\quad
    a_{2,2}^{(2)}=a_{2,2}^{(3)}= \frac{2N(N+2)^2}{9}\,, \quad  
     a_{2,2}^{(4)}= \frac{4N(N+2)}{3},\\
     &a_{2,2}^{(5)}= \frac{2N(N+2)}{9}, \quad
  a_{2,2}^{(6)}= \frac{4N(N+2)}{9}, \quad 
  a_{2,2}^{(7)}= \frac{N(N+2)^2}{9}, \quad 
  a_{2,2}^{(8)}= a_{2,2}^{(9)}=\frac{2N(N+2)}{3}\,. \nonumber
\end{align}
Lastly, we denote the order $\lambda^3_0 h_0$ contribution to the one-point function \eqref{1ptordeps3} by $\CJ_{3,1}$ which can be determined from the component of the two-point function proportional to $\lambda_0^3$:
\begin{gather}
   h_0 \lambda^3_0\CJ_{3,1}= -h_0\int d^2X_2 \langle \phi_0^2(X_1)\phi_0 ^2(X_2) \rangle_{4-\text{loop part }}\,.
   \label{CJ3,1}
\end{gather}
Since we are only interested in the divergent part of $\CJ_{3,1}$, we can infer it directly from the wavefunction renormalization $Z_{\phi^2}$ by requiring the finiteness of $Z^{-2}_{\phi^2}\langle \phi_0^2(X_1)\phi_0^2(X_2)\rangle$ at order $\lambda^3$. The result is given below, 
\begin{equation}
    \begin{split}
    \CJ_{3,1} &\to\frac{  N (N+2)}{\eta^{2-5\epsilon}}\bigg(\frac{ (N+4) (N+5)}{221184 \pi ^9 \epsilon^3}-\frac{12 N^2+91 N+164-3 \chi  (N+4) (N+5)}{1327104 \pi ^9 \epsilon^2}\nonumber\\
    &+\frac{5 \pi ^2 N (N+9)+32 N (6 N+47)}{10616832 \pi ^9 \epsilon}\nonumber\\
    &+\frac{4 \left(-164 \chi +25 \pi ^2+734\right)+6 \chi ^2 (N+4) (N+5)-4 \chi  N (12 N+91)}{10616832 \pi ^9 \epsilon}\bigg)\,,
\end{split}
\end{equation}
where 
\begin{gather}
    \chi=12+5 \log \left(\pi  e^{\gamma_E }\right)\,.
\end{gather}

The diagrams in Figure~\ref{4loopdiag} are computed using the methods outlined in Section \ref{MassOperator}.  Here we present the final results for the divergent parts of these integrals, and provide additional details for $\CJ_{1,3}^{(2)}$ and $\CJ_{2,2}^{(9)}$, as their computations are more involved.
The first diagram in Figure~\ref{4loopdiag} produces the following divergent contribution as $\ep \to 0$,
\begin{align}
    \CJ_{0,4}^{}&=C_\phi^5\int \frac{d^2 X_1 d^2 X_2 d^2 X_3 d^2 X_4}{(x^2_{1}+\eta^2)^{\frac{d-2}{2}}X_{12}^{d-2}X_{23}^{d-2}X_{34}^{d-2}(x^2_{4}+\eta^2)^{\frac{d-2}{2}}}
\nonumber\\
     &\to\frac{48+\left(6 (\chi -8) \chi +5 \pi ^2+480\right) \epsilon ^2+24 (\chi -4) \epsilon }{6144 \pi ^6 \epsilon ^3 \eta^{2-5\epsilon}}\,.
\end{align}
For diagrams $\CJ_{1,3}^{(1)}$ to $\CJ_{1,3}^{(4)}$ in Figure~\ref{4loopdiag}, we denote the bulk integration as $X_0$ and points on the defect as $X_1=(x_1, 0)$, $X_2=(x_2, 0)$, $X_3=(x_3, 0)$ and we also set $X_4 = (0, \eta)$.  For example we find the divergent part of $\CJ_{1,3}^{(1)}$ below,
\begin{align}
     \CJ_{1,3}^{(1)}=\int \frac{C_\phi^6d^2 X_1 d^2 X_2 d^2 X_3 d^d X_0}{X_{04}^{2(d-2)}X_{01}^{d-2}X_{03}^{d-2}X_{12}^{d-2}X_{23}^{d-2}}
     \to\frac{48+\left(6 (\chi -8) \chi +5 \pi ^2+336\right) \epsilon ^2+24 (\chi -4) \epsilon }{18432 \pi ^7 \epsilon ^3\eta^{2-5\epsilon}}\,.
\end{align}
The third diagram in Figure \ref{4loopdiag} can be first simplified by taking the integral over $X_1,X_2$ and representing the bulk point as $X_{0}=(x_0,\eta_0)$,
\begin{gather}
    \CJ_{1,3}^{(2)}=C_\phi^6\int \frac{d^2 X_1 d^2 X_2 d^2 X_3 d^d X_0}{X_{01}^{d-2}X_{02}^{d-2}X_{03}^{d-2}X_{04}^{d-2}X_{12}^{d-2}X_{34}^{d-2}}\\
    =C_\phi^6 \Gamma_{0}\left(2;\frac{d-2}{2},\frac{d-2}{2},\frac{d-2}{2}\right)\int \frac{ d^2 x_3 d^2 x_0d^{d-2}\eta_{0}}{\left(x^2_{03}+\eta^2_{0}\right)^{\frac{d-2}{2}}\left(x^2_0+(\eta-\eta_0)^2\right)^{\frac{d-2}{2}}\left(x^2_{3}+\eta^2\right)^{\frac{d-2}{2}}\eta^{3d-10}_{0}}\,.\nonumber
\end{gather}
Next we use the Mellin-Barnes representation for
\begin{gather}
    \frac{1}{\left(x^2_{03}+\eta^2_0\right)^{\frac{d-2}{2}}}=\int\limits_{-i\infty}^{+i\infty}\frac{dz_1}{2\pi i}\frac{\Gamma(-z_1)\Gamma(z_1+\frac{d-2}{2})}{\Gamma(\frac{d-2}{2})}\frac{1}{x^{-2z_1}_{03}\eta^{2z_1+d-2}_0}\,,\nonumber\\
    \frac{1}{\left(x^2_{0}+(\eta_0-\eta)^2\right)^{\frac{d-2}{2}}}=\int\limits_{-i\infty}^{+i\infty}\frac{dz_2}{2\pi i}\frac{\Gamma(-z_2)\Gamma(z_2+\frac{d-2}{2})}{\Gamma(\frac{d-2}{2})}\frac{1}{x^{-2z_2}_{0}(\eta_0-\eta)^{2z_2+d-2}}\,.
\end{gather}
By integrating over the bulk point using \eqref{Cdef}, the computation simplifies to a trivial integral over $x_3$, yielding the following result,
\begin{gather}
\CJ_{1,3}^{(2)}
    =\frac{C_\phi^6 }{\eta^{5d-18}}\Gamma_{0}\left(2;\frac{d-2}{2},\frac{d-2}{2},\frac{d-2}{2}\right)\int\limits_{-i\infty}^{+i\infty}\frac{dz_1dz_2}{(2\pi i)^2}\prod\limits_{i=1}^{2}\left(\frac{\Gamma(-z_i)\Gamma(z_i+\frac{d-2}{2})}{\Gamma(\frac{d-2}{2})}\right)\nonumber\\
    \times C_{2;-z_1,-z_2}C_{d-2;z_1+2d-6,z_2+\frac{d-2}{2}} 
\frac{\pi \Gamma(2+z_1+z_2)\Gamma(\frac{d}{2}-3-z_1-z_2)}{\Gamma(\frac{d-2}{2})}\,.
\end{gather}
\normalsize
For this double MB integral, the $\epsilon$-expansion generated by the \texttt{MB.m} program is the sum of $$\frac{\epsilon \left(6 \epsilon \left(\chi ^2+80\right)-23 \pi ^2 \epsilon+24 \chi \right)+48}{2^{14}\times 3^2 \pi ^7 \epsilon^3}$$ and two contour integrals at the order $\frac{1}{\epsilon}$. The integrands of these two contour integral can be unified by appropriately shifting the contour and the combined integral reduces to,
\begin{align}
   \frac{1}{\epsilon}\int_{-\frac{1}{2}+i\mathbb R}\frac{dz}{2\pi i} \frac{\csc ^2(\pi  z) \left(13 \pi ^2 z^2 \csc ^2(\pi  z)-12 z^2 \psi ^{(1)}(z)+12\right)}{24576 \pi ^5
   z^2}\,,
\end{align}
which includes three independent parts
\begin{align}
    I_1&= \int_{-\frac{1}{2}+i\mathbb R}\frac{dz}{2\pi i}\frac{\pi^2}{\sin^4(\pi z)} = \frac{2}{3}\,, \nonumber\\
    I_2& = \int_{-\frac{1}{2}+i\mathbb R}\frac{dz}{2\pi i} \frac{\pi^2}{z^2\sin^2(\pi z)}= 2\zeta(3)\,,\\
     I_3&= \int_{-\frac{1}{2}+i\mathbb R}\frac{dz}{2\pi i} \frac{\pi^2\psi^{(1)}(z)}{\sin(\pi z)^2}= 2 \zeta(3)+\frac{\pi^2}{3}\,.\nonumber
\end{align}
The integral $I_2$ can be evaluated by closing the contour on the right and summing up the residues. For  $I_3$, we move the contour to Re$(z) =\frac{1}{2}$,
\begin{align}
    I_3 = 2 \zeta(3)+\int_{\frac{1}{2}+i\mathbb R}\frac{dz}{2\pi i} \frac{\pi^2\psi^{(1)}(z)}{\sin(\pi z)^2}\,,
    \label{I3int}
\end{align}
and make use of the following integral representation (which is convergent)
\begin{align}\label{pisint}
    \psi^{(1)}(z) = \int_0^\infty dt \frac{t\, e^{-z t}}{1-e^{-t}}\,.
\end{align}
We evaluate the $z$ integral in \eqref{I3int} by closing the contour and then perform the remaining $t$ integral to obtain,
\begin{align}
    I_3 = 2 \zeta(3)+\int_0^\infty  dt\,\frac{ t^2 e^{-t}}{(1-e^{-t})^2} = 2 \zeta(3)+\frac{\pi^2}{3}\,.
\end{align}
Putting everything together we deduce the divergent part of $\CJ_{1,3}^{(2)}$ below,
\begin{gather}
\CJ_{1,3}^{(2)}
    \to \frac{1}{\eta^{2-5\epsilon}}\frac{48+\epsilon  \left(24\chi +6 \chi ^2 \epsilon +5 \left(96+\pi ^2\right) \epsilon \right)}{147456 \pi ^7 \epsilon ^3}\,.
\end{gather}
By similar calculations, we also have
\ie 
    \CJ_{1,3}^{(3)}=&\int \frac{C_\phi^6d^2 X_1 d^2 X_2 d^2 X_3 d^d X_0}{X_{01}^{2(d-2)}X_{02}^{d-2}X_{04}^{d-2}X_{23}^{d-2}X_{34}^{d-2}}
    \to \frac{48+\left(6 (\chi -4) \chi +5 \pi ^2+432\right) \epsilon ^2+24 (\chi -2) \epsilon }{49152 \pi ^7 \epsilon ^3 \eta^{2-5\epsilon}}\,,
\\
    \CJ_{1,3}^{(4)}=&\int \frac{C_\phi^6 d^2 X_1 d^2 X_2 d^2 X_3 d^d X_0}{X_{01}^{d-2}X_{02}^{2(d-2)}X_{03}^{d-2}X_{14}^{d-2}X_{34}^{d-2}}
   \to\frac{48+\left(6 (\chi -4) \chi +5 \pi ^2+432\right) \epsilon ^2+24 (\chi -2) \epsilon}{73728 \pi ^7 \epsilon ^3 \eta^{2-5\epsilon}}\,.
\fe  
For diagrams $\CJ_{2,2}^{(1)}$ to $\CJ_{2,2}^{(9)}$ in Figure~\ref{4loopdiag}, we denote the bulk points that are being integrated as $X_0,X_3$ and points on the defect as $X_1=(x_1, 0)$, $X_2=(x_2, 0)$ and we set $X_4 = (0, \eta)$. The  divergent parts of the first eight integrals are summarized below,
\ie 
    &\CJ^{(1)}_{2, 2} = \int \frac{C_\phi^7d^2 X_1 d^2 X_2 d^d X_0 d^dX_{3}}{X_{01}^{d-2}X_{02}^{d-2}X_{03}^{2(d-2)}X_{12}^{d-2}X_{34}^{2(d-2)}}
     \to\frac{48+\left(6 (\chi -8) \chi +5 \pi ^2+240\right) \epsilon ^2+24 (\chi -4) \epsilon }{49152 \pi ^8 \epsilon ^3 \eta^{2-5\epsilon}}\,.
\\&
    \CJ^{(2)}_{2, 2} = \int \frac{C_\phi^7d^2 X_1 d^2 X_2 d^d X_0 d^dX_{3}}{X_{01}^{2(d-2)}X_{02}^{d-2}X_{03}^{d-2}X_{23}^{d-2}X_{34}^{2(d-2)}}
     \to\frac{48+\left(6 (\chi -4) \chi +5 \pi ^2+288\right) \epsilon ^2+24 (\chi -2) \epsilon}{147456 \pi ^8 \epsilon ^3 \eta^{2-5\epsilon}}\,.
\\&
    \CJ^{(3)}_{2, 2} = \int \frac{C_\phi^7d^2 X_1 d^2 X_2 d^d X_0 d^dX_{3}}{X_{01}^{2(d-2)}X_{03}^{2(d-2)}X_{23}^{d-2}X_{24}^{d-2}X_{34}^{d-2}}
     \to\frac{48+\epsilon  \left(24 \chi +6 \left(\chi ^2+72\right) \epsilon +5 \pi ^2 \epsilon \right)}{294912 \pi ^8 \epsilon ^3\eta^{2-5\epsilon}}\,.
\\&
    \CJ^{(4)}_{2, 2} = \int \frac{C_\phi^7d^2 X_1 d^2 X_2 d^d X_0 d^dX_{3}}{X_{01}^{d-2}X_{02}^{d-2}X_{03}^{2(d-2)}X_{13}^{d-2}X_{24}^{d-2}X_{34}^{d-2}}
     \to\frac{\left(462+5 \pi ^2\right) \epsilon ^2+6 ((\chi +1) \epsilon +2)^2+24}{589824 \pi ^8 \epsilon ^3 \eta^{2-5\epsilon}}\,.
\\&
    \CJ^{(5)}_{2, 2} = C_\phi^7\int \frac{d^2 X_1 d^2 X_2 d^d X_0 d^dX_{3}}{X_{01}^{d-2}X_{03}^{3(d-2)}X_{23}^{d-2}X_{14}^{d-2}X_{24}^{d-2}}
     \to-\frac{(2 \chi  +5) \epsilon +4}{196608 \pi ^8 \epsilon ^2 \eta^{2-5\epsilon}}\,.
\\&
    \CJ^{(6)}_{2, 2} = C_\phi^7\int \frac{d^2 X_1 d^2 X_2 d^d X_0 d^dX_{3}}{X_{01}^{d-2}X_{03}^{3(d-2)}X_{12}^{d-2}X_{24}^{d-2}X_{34}^{d-2}}
     \to-\frac{(2 \chi  +1)\epsilon +4}{65536 \pi ^8 \epsilon ^2 \eta^{2-5\epsilon}}\,.
\\&
    \CJ^{(7)}_{2, 2} = C_\phi^7\int \frac{d^2 X_1 d^2 X_2 d^d X_0 d^dX_{3}}{X_{01}^{2(d-2)}X_{03}^{d-2}X_{04}^{d-2}X_{23}^{2(d-2)}X_{34}^{d-2}}
     \to\frac{48+\epsilon  \left(24 \chi +6 \left(\chi ^2+72\right) \epsilon +5 \pi ^2 \epsilon \right)}{294912 \pi ^8 \epsilon ^3 \eta^{2-5\epsilon}}\,.
\\&
     \CJ^{(8)}_{2, 2}=\int \frac{C_\phi^7 d^d X_0 d^dX_{3}d^2 X_1 d^2 X_2}{X_{01}^{d-2}X_{03}^{2(d-2)}X_{04}^{d-2}X_{12}^{d-2}X_{23}^{d-2}X_{34}^{d-2}} \to\frac{240\!+\!\left(30 (\chi \!-\!6) \chi \!+\!25 \pi ^2\!+\!1044\right) \epsilon ^2\!+\!120 (\chi \!-\!3) \epsilon }{589824 \pi ^8 \epsilon ^3 \eta^{2-5\epsilon}}\,.
\fe 
The integral for the last diagram in Figure~\ref{4loopdiag} is
\begin{gather}
     \CJ^{(9)}_{2, 2}=C_\phi^7\int \frac{ d^d X_0 d^dX_{3}d^2 X_1 d^2 X_2}{X_{01}^{d-2}X_{02}^{d-2}X_{03}^{d-2}X_{04}^{d-2}X_{13}^{d-2}X_{23}^{d-2}X_{34}^{d-2}}\,.
\end{gather}
To evaluate this, we first integrate over the bulk point $X_0$ using Feynman parametrization,
\begin{gather}
    \int \frac{ d^d X_0 }{X_{01}^{2\alpha_1}X_{02}^{2\alpha_2}X_{03}^{2\alpha_3}X_{04}^{2\alpha_4}}=\frac{\Gamma(\alpha)}{\prod \limits_{i=1}^{4}\Gamma(\alpha_i)}\int\limits_{0}^{1}\prod_{a=1}^{4}du_a \delta\left(\sum\limits_{c=1}^{4}u_c-1\right)\int d^dX_0\frac{u^{\alpha_1-1}_1u^{\alpha_2-1}_2u^{\alpha_3-1}_3u^{\alpha_4-1}_4}{\left(\sum\limits_{i=1}^{4}u_i X^2_{0 i}\right)^\alpha}\nonumber\\
    =\pi^{\frac{d}{2}}\frac{\Gamma(\alpha-\frac{d}{2})}{\prod \limits_{i=1}^{4}\Gamma(\alpha_i)}\int\limits_{0}^{1}\prod_{a=1}^{4}du_a \delta\left(\sum\limits_{c=1}^{4}u_c-1\right)\frac{u^{\alpha_1-1}_1u^{\alpha_2-1}_2u^{\alpha_3-1}_3u^{\alpha_4-1}_4}{\left(\sum\limits_{i<j}^{4}u_i u_j X^2_{ ij}\right)^{\alpha-\frac{d}{2}}}
\end{gather}
where $\alpha=\sum \limits_{i=1}^{4}\alpha_i$. Next we introduce a five-dimensional Mellin-Barnes representation using \eqref{basicMB} and then perform the integration over the Feynman parameters using \eqref{us},
\begin{gather}
     \CJ^{(9)}_{2, 2}=C_\phi^7\int\limits_{-i\infty}^{+i\infty} \prod\limits_{p=1}^{5}\frac{dz_p}{2\pi i}\pi^{\frac{d}{2}}\frac{\Gamma(Z+\alpha-\frac{d}{2})}{\prod \limits_{i=1}^{4}\Gamma(\alpha_i)}\frac{\prod \limits_{j=1}^{5}\Gamma(-z_j)\prod\limits_{k=1}^{4}\Gamma(b_k)}{\Gamma(b_1+b_2+b_3+b_4)}\nonumber\\
     \times\int \frac{d^dX_{3}d^2 X_1 d^2 X_2}{X_{13}^{2(-z_2+\frac{d-2}{2})}X_{23}^{2(-z_4+\frac{d-2}{2})}X_{34}^{2(-z_1+\frac{d-2}{2})}X_{12}^{2(Z+\alpha-\frac{d}{2})}X_{14}^{-2z_3}X_{24}^{-2z_5}}\,,
\end{gather}
where $Z=\sum \limits_{i=1}^{5}z_i$ and 
\begin{gather}
    b_1=\alpha_1+z_2+z_3-Z-\alpha+\frac{d}{2}\,,\quad
    b_2=\alpha_2+z_4+z_5-Z-\alpha+\frac{d}{2}\,,\nonumber\\
    b_3=\alpha_3+z_1+z_2+z_4\,,\quad
    b_4=\alpha_4+z_1+z_3+z_5\,.
\end{gather}
Finally we integrate over the bulk point $X_3$ using \eqref{Sdef0}, apply \eqref{G0int} for remaining $X_1,X_2$ integration, and obtain the divergent part from the MB integral below,
\begin{align}
    \CJ^{(9)}_{2, 2}
    &=C_\phi^7\int\limits_{-i\infty}^{+i\infty} \prod\limits_{p=1}^{5}\frac{dz_p}{2\pi i}\pi^{\frac{d}{2}}\frac{\Gamma(Z+\alpha-\frac{d}{2})}{\prod \limits_{i=1}^{4}\Gamma(\alpha_i)}\frac{\prod \limits_{j=1}^{5}\Gamma(-z_j)\prod\limits_{k=1}^{4}\Gamma(b_k)}{\Gamma(b_1+b_2+b_3+b_4)}\nonumber\\
   & \times S_{d}\left(-z_2+\frac{d-2}{2},-z_4+\frac{d-2}{2},-z_1+\frac{d-2}{2},z_6,z_7\right)\\
   &\times \Gamma_{0}\left(2;-z_3-z_6,-z_5-z_7,z_3+z_5+z_6+z_7+\alpha+\frac{d-6}{2}\right)
     \to\frac{\zeta (3)}{2^{12}\pi ^8 \epsilon \eta^{2-5\epsilon} }\,.\nonumber
\end{align}
By combining the above results in \eqref{1ptordeps3} and requiring the one-point function $\left\langle \left[\phi^2\right]_{R}\right\rangle$ to remain finite in the $\ep\to 0$ limit, 
we obtain the following renormalization of $h_0$,
\begin{align}\label{hh0}
h_0 &= \mu^\epsilon\Bigg[h+\frac{h^2}{\pi  \epsilon }+\frac{h^3}{\pi ^2 \epsilon ^2}+\frac{h^4}{\pi ^3 \epsilon ^3}\nonumber\\
&+\frac{(N+2)\lambda h}{3(4\pi)^2\epsilon}-\frac{N+2}{3}\left(\frac{h^2 \lambda }{16 \pi ^3}+\frac{5 h \lambda ^2}{3\times 2^{10}\pi ^4}\right)\frac{1}{\epsilon}+\frac{N+2}{3}\left(\frac{ 3 h^2 \lambda }{2^5 \pi ^3}+\frac{(N+5)h \lambda ^2}{3\times 2^8 \pi ^4}\right)\frac{1}{\epsilon^2}\nonumber\\
&+h \lambda^3 \left(\frac{(N+2) (N+5) (N+6)}{27 (4 \pi )^6 \epsilon ^3}-\frac{(N+2) (61 N+278)}{324 (4 \pi )^6 \epsilon ^2}+\frac{(N+2) (5 N+37)}{108 (4 \pi )^6 \epsilon }\right)\nonumber\\
&+h^2 \lambda ^2 \bigg(
\frac{(N+2) (11 N+46)}{432 (2 \pi )^5 \epsilon ^3}-\frac{(N+2) (20 N+89)}{27 (4 \pi )^5 \epsilon ^2}+\frac{(N+2) (16 N+144 \zeta (3)+151)}{108 (4 \pi )^5 \epsilon}\bigg)\nonumber\\
&+h^3 \lambda \frac{N+2}{3} \left(\frac{1}{8 \pi ^4 \epsilon ^3}-\frac{1}{8 \pi ^4 \epsilon ^2}+\frac{1}{24 \pi ^4 \epsilon}\right)\Bigg]\,,
\end{align}
which is valid to the fourth order of the couplings.
Imposing $\mu\partial_\mu h_0=0$ and using $\beta_\lambda$ from \eqref{betalambda}, we obtain the beta function of $h$ to the fourth order in coupling constants,
\begin{gather}\label{betah4}
    \beta_h = -\epsilon h+\frac{h^2}{\pi }+\frac{N+2}{3}\left(\frac{h \lambda }{(4 \pi) ^2}-\frac{h^2 \lambda }{(2 \pi) ^3}-\frac{5 h \lambda ^2}{6(4 \pi) ^4}\right)+h \lambda ^3 \frac{(N+2) (5 N+37)}{36 (4 \pi )^6}\nonumber\\+h^2 \lambda^2
    \frac{ (N+2) (16 N+144 \zeta (3)+151)}{36 (4 \pi )^5}+\frac{2 h^3 \lambda  (N+2)}{3 (2 \pi )^4}\,.
\end{gather}
Plugging in the bulk fixed point in $\lambda_\star$ from \eqref{fixedlambda} into $\beta_h = 0$ yields the defect fixed point,
\begin{align}\label{hfixedpteps3}
h_\star &= \frac{6 \pi  \epsilon }{N+8}+\frac{(N+2)(11N+148) \pi  \epsilon ^2}{2(N+8)^3}\nonumber\\
&+\frac{  (N+2)  \left(43 N^3+679 N^2+48 (N-28) (N+8) \zeta (3)+1984 N-19968\right)}{8 (N+8)^5}\pi\epsilon ^3+\cO(\ep^4)\,.
\end{align}
We can also extract from $\B_h$ the $\ep^3$ correction to the scaling dimension \eqref{Dphi2} of the  defect operator   $[\hat\phi^2]_R$,
\ie \label{Dphi2ordeps3}
&\left. \Delta_{\hat\phi^2} \right|_{\ep^3}  =   72\frac{N+2}{(N+8)^3}\epsilon^3 
\\
&-\frac{-3 N^4+446 N^3+3576 N^2+10656 N+10624-96 (N+2) (N+8) (5 N+22) \zeta (3)}{8 (N+8)^5}\epsilon^3\,.
\fe 
The results here also determine the one-point function coefficient of $\phi^2$ in \eqref{ta2} to the next order,
\ie \label{ta3} 
\left. a_{\phi^2} \right|_{\ep^3} &= 
\frac{N \epsilon ^3 }{64 \pi ^2 (N+8)^5}\bigg(N^4+N^3 \left(-48 \zeta (3)-6 \pi ^2+361+168 \log \left(\pi e^{\gamma_E} \right)\right)\\
&-18 N^2 \left(-48 \zeta (3)+8 \pi ^2-223+4 \log \left(\pi e^{\gamma_E} \right) \left(3 \log \left(\pi e^{\gamma_E} \right)-22\right)\right)\\
&-64 N \left(-198 \zeta (3)+18 \pi ^2-389+27 \log \left(\pi e^{\gamma_E}\right) \left(2 \log \left(\pi e^{\gamma_E} \right)-1\right)\right)\\
&-256 \left(-84 \zeta (3)+12 \pi ^2-143+6 \log \left(\pi e^{\gamma_E}\right) \left(1+9 \log \left(\pi e^{\gamma_E}\right)\right)\right)\bigg) \,,
\fe 
and similarly for the normalized mass operator $\cO_2$,
\ie \label{a2eps3}
\left.a_{\CO_2}\right|_{\ep^3} =\frac{\sqrt{N} (N+2) \left(15 N^3+196 N^2+48 (N-28) (N+8) \zeta (3)+1072 N-10688\right)}{16 \sqrt{2} (N+8)^5} \epsilon ^3\,.
\fe 
As a consistency check, we find that the leading large $N$ behavior 
\ie 
a_{\CO_2} =\frac{1}{\sqrt{2N}}\left(3\epsilon+\frac{5}{4}\epsilon^2+\frac{15 \epsilon ^3}{16}\right)+\cdots 
\fe 
agrees with  \cite{Giombi:2023dqs} up to an overall minus sign.

\section{Consistency check of the ``shadow relation''} \label{checkSurp}
In this appendix, we provide an additional consistency check for the ``shadow relation'' based on general considerations. Specifically, without performing any diagrammatic computations, we start with the following ansatz for bare coupling in terms of the renormalized couplings:
\begin{align}\label{h0bareansatz}
h_0 &= \mu^\epsilon\Bigg[h+\frac{h^2}{\pi  \epsilon }+\frac{h^3}{\pi ^2 \epsilon ^2}+\frac{h^4}{\pi ^3 \epsilon ^3}\nonumber\\
&+\frac{(N+2)\lambda h}{3(4\pi)^2\epsilon}-\frac{N+2}{3}\left(\frac{h^2 \lambda }{16 \pi ^3}+\frac{5 h \lambda ^2}{3\times 2^{10}\pi ^4}\right)\frac{1}{\epsilon}+\frac{N+2}{3}\left(\frac{ 3 h^2 \lambda }{2^5 \pi ^3}+\frac{(N+5)h \lambda ^2}{3\times 2^8 \pi ^4}\right)\frac{1}{\epsilon^2}\nonumber\\
&+h \lambda^3 \left(\frac{(N+2) (N+5) (N+6)}{27 (4 \pi )^6 \epsilon ^3}-\frac{(N+2) (61 N+278)}{324 (4 \pi )^6 \epsilon ^2}+\frac{(N+2) (5 N+37)}{108 (4 \pi )^6 \epsilon }\right)\nonumber\\
&+h^2 \lambda^2 \bigg(\frac{q_3}{\epsilon^3}+\frac{q_2}{\epsilon^2}+\frac{q_1}{\epsilon}\bigg)+h^3 \lambda \bigg(\frac{r_3}{\epsilon^3}+\frac{r_2}{\epsilon^2}+\frac{r_1}{\epsilon}\bigg)\Bigg]\,,
\end{align}
where terms proportional to $h\lambda^3$ were determined using the four-loop renormalization of the bulk two-point function $\langle \phi_0^2\phi_0^2\rangle$ (see around \eqref{CJ3,1}). Here $q_{1,2,3}$ and $r_{1,2,3}$ are constant coefficients.

Requiring that the beta function derived from \eqref{h0bareansatz} has no divergences as $\ep\to 0$ up to fourth order in the coupling constants allows us to fix the coefficients  $q_2, q_3$ as well as $r_2,r_3$. This reproduces the results obtained from diagrammatic computations in \eqref{hh0}.

Substituting the bulk fixed point value $\lambda_\star$ from \eqref{fixedlambda} into the condition $\beta_h = 0$ yields the defect fixed point as a function of $q_1$ and $r_1$,
\begin{gather}
    h_\star= \frac{6 \pi  \epsilon }{N+8}+\frac{(N+2)(11N+148) \pi  \epsilon ^2}{2(N+8)^3}-\frac{5184 \pi ^5  (8 \pi  q_1+r_1)}{(N+8)^3}\epsilon ^3\nonumber\\
    +\frac{ (N+2)  \left(91 N^3+2476 N^2+21520 N+45888+96 (N+8) (5 N+22) \zeta (3)\right)}{8 (N+8)^5}\pi \epsilon ^3\,.
\end{gather}
We then calculate the scaling dimension of $\hat{\phi}^2$ from the derivative of the beta function. Combining it with the bulk counterpart \eqref{Dbulkphi211} gives the following identity,
\begin{gather}
   \Delta_{\phi^2} + \Delta_{\hat\phi^2} = 4 + \frac{5184 \pi ^4 \epsilon^3}{(N+8)^3}r_1+\CO(\epsilon^4)\,,
\end{gather}
indicating that only contributions from $\CJ_{1,3}^{(1)}-\CJ_{1,3}^{(4)}$  (terms of order $h^3 \lambda$) are relevant to this relation. Finally, for $r_1$ given in \eqref{hh0} we reproduce \eqref{surp}.

\bibliography{draft}
\bibliographystyle{utphys}

\end{document}